\documentclass[amsmath, pra, twocolumn, showpacs, floatfix]{revtex4-1}
\usepackage{graphicx}
\usepackage{dsfont}

\begin{document}   

\title{Channeling of spontaneous emission from an atom into the fundamental and higher-order modes of a vacuum-clad ultrathin optical fiber}
 
\author{Fam Le Kien}
\affiliation{Quantum Systems Unit, Okinawa Institute of Science and Technology Graduate University, Onna, Okinawa 904-0495, Japan}

\author{S. Sahar S. Hejazi}
\affiliation{Quantum Systems Unit, Okinawa Institute of Science and Technology Graduate University, Onna, Okinawa 904-0495, Japan}

\author{Thomas Busch}
\affiliation{Quantum Systems Unit, Okinawa Institute of Science and Technology Graduate University, Onna, Okinawa 904-0495, Japan}

\author{Viet Giang Truong}
\affiliation{Light-Matter Interactions Unit, Okinawa Institute of Science and Technology Graduate University, Onna, Okinawa 904-0495, Japan}

\author{S\'{i}le Nic Chormaic}
\affiliation{Light-Matter Interactions Unit, Okinawa Institute of Science and Technology Graduate University, Onna, Okinawa 904-0495, Japan}
\affiliation{School of Chemistry and Physics, University of KwaZulu-Natal, Durban, KwaZulu-Natal, 4001, South Africa}

\date{\today}

\begin{abstract}

We study spontaneous emission from a rubidium atom into the fundamental and higher-order modes of a vacuum-clad ultrathin optical fiber.
We show that the spontaneous emission rate depends on the magnetic sublevel, the type of modes, the orientation of the quantization axis, and the fiber radius. We find that the rate of spontaneous emission into the TE modes is always symmetric with respect to the propagation directions. 
Directional asymmetry of spontaneous emission into other modes may appear when the quantization axis does not lie in the meridional plane containing the position of the atom. When the fiber radius is in the range from 330 nm to 450 nm, the spontaneous emission into the HE$_{21}$ modes is stronger than into the HE$_{11}$, TE$_{01}$, and TM$_{01}$ modes. At the cutoff for higher-order modes, the rates of spontaneous emission into guided and radiation modes undergo steep variations, which are caused by the changes in the mode structure. We show that the spontaneous emission from the upper level of the cyclic transition into the TM modes is unidirectional when the quantization axis lies at an appropriate azimuthal angle in the fiber transverse plane. 

\end{abstract}

\maketitle

\section{Introduction}

Optical fibers can be tapered to a diameter comparable to or smaller than the wavelength of light \cite{Mazur's Nature,Birks,taper}. 
Due to the tapering, the original core almost vanishes and the refractive indices that determine the guiding properties of the tapered fiber are those of the original silica cladding and the surrounding vacuum. Since the diameters of such tapered fibers are in the range of a few hundred nanometers, they are usually called nanofibers. When the radius of the fiber is small enough, it can support only a single mode in the optical region of frequency.

In a vacuum-clad nanofiber, the guided field penetrates an appreciable distance into the surrounding medium and appears as an evanescent wave carrying a significant fraction of the power and having a complex polarization pattern \cite{Bures99,Tong04,fibermode}. 
Nanofibers are therefore versatile tools for coupling light and matter and have a wide range of potential practical applications \cite{Morrissey13,Chormaic2016c}. For example, they have been used for trapping atoms \cite{fiber trap,Vetsch10,Goban12}, for probing atoms \cite{Domokos02,absorption,Nayak07,Nayak09,Sile2009,Dawkins11,Reitz13,Russell13,Sile2016}, molecules \cite{Stiebeiner09}, quantum dots \cite{Yalla12}, and color centers in nanodiamonds \cite{Schroder12,Liebermeister13}, and for mechanical manipulation of small particles \cite{Skelton12,Brambilla07,Fam2013}. 

Tapered fibers can also be fabricated with slightly larger diameters or larger refractive indices so that they can support not only the fundamental HE$_{11}$ mode but also several higher-order modes. Compared to the HE$_{11}$ mode, the higher-order modes have larger cutoff size parameters and more complex intensity, phase, and polarization distributions. For ease of reference, the vacuum-clad tapered fibers that can support the fundamental mode and several higher-order modes are called ultrathin optical fibers in this paper. 

It has been shown that ultrathin optical fibers with higher-order modes can be used to trap, probe, and manipulate atoms, molecules, and particles \cite{Tong2007,Tong2008,Rauschenbeutel2008,Minogin2013,Busch2013,Chormaic2016a,Reinhard2016}. 
The excitation of higher-order modes has been studied \cite{Volpe2004,Chormaic2011}.
The production of ultrathin fibers with higher-order modes \cite{Chormaic2012,Fatemi2013,Chormaic2014}
and the experimental studies on the interaction with atoms \cite{Chormaic2015a} or particles \cite{Chormaic2015b,Chormaic2016b} have been reported.
The possibility to control and manipulate individual atoms near an ultrathin fiber can also find applications for quantum information. 

The interaction between guided light and atoms is of academic and practical interest.
Many applications require a deep understanding and an effective control of spontaneous emission of atoms near an ultrathin optical fiber. 
Radiative decay of an atom in the vicinity of a nanofiber has been studied in the context of a two-level atom \cite{Jhe,Tromborg,Klimov} as well as a realistic multilevel atom with a hyperfine structure of energy levels \cite{cesium decay,Fam2008}.
The parameters for the decay of populations \cite{Jhe,Tromborg,Klimov,cesium decay,Fam2008} and cross-level coherences \cite{cesium decay,Fam2008,ChangMinogin} have been calculated. 

Recently, emission of particles with circularly polarized dipoles began to attract much attention \cite{Lee2012,Lin2013,Mueller2013,Zayats2013,Ming2013,Leuchs2014,Banzer15,Zayats,Dogariu}.
It has been shown that the near-field interference of a circularly polarized dipole coupled to a dielectric or metallic object leads to unidirectional excitation of guided modes or surface plasmon polariton modes \cite{Lee2012,Lin2013,Mueller2013,Zayats2013,Ming2013,Leuchs2014,Banzer15}. This effect has been experimentally demonstrated by shining circularly polarized light onto a nanoslit \cite{Lee2012,Zayats2013} or closely spaced subwavelength apertures \cite{Lin2013} in a metal film and by exciting a nanoparticle on a dielectric interface with a tightly focused vector light beam \cite{Leuchs2014,Banzer15}. 

It has been shown that spontaneous emission and scattering from an atom with a circular dipole near a nanofiber can be asymmetric with respect to the opposite axial propagation directions \cite{Mitsch14b,Petersen2014,Fam2014,AtomArray,Scheel15,Sayrin15b}. These directional effects are the signatures of spin-orbit coupling of light \cite{Zeldovich,Bliokh review,Bliokh review2015,Bliokh2014,Bliokh2015} carrying transverse spin angular momentum \cite{Bliokh2014,Banzer review2015}. They are due to the existence of a nonzero longitudinal component of the nanofiber guided field, which oscillates in phase quadrature with respect to the radial transverse component. The possibility of directional emission from an atom into propagating radiation modes of a nanofiber and the possibility of generation of a lateral force on the atom have been reported \cite{Scheel15}. The direction-dependent emission and absorption of photons lead to chiral quantum optics \cite{Lodahl2017}. 

Spontaneous emission from a multilevel atom into the fundamental and higher-order modes of an ultrathin fiber has been studied
by Masalov and Minogin \cite{Minogin2014}. They have found that the decay rates into the higher-order modes can be significantly larger than into
the fundamental mode. Their calculations were limited to single transitions and single polarizations. 
However, all types of transitions and polarizations must be accounted for in a realistic situation. 
In addition, in Ref.~\cite{Minogin2014} the fiber axis was used as the quantization axis and consequently no direction dependencies of the rates could be observed. Moreover, emission into radiation modes was not considered in Ref.~\cite{Minogin2014}.

The aim of the present paper is to investigate directional spontaneous emission from a multilevel atom with an arbitrary quantization axis into an ultrathin fiber. We calculate the rates of spontaneous emission into the fundamental and higher-order guided modes propagating in a given direction. We also calculate the rate of spontaneous emission into radiation modes.

The paper is organized as follows. In Sec.~\ref{sec:model} we describe the interaction of an alkali-metal atom with the electromagnetic field in the presence
of an ultrathin optical fiber.
Section \ref{sec:spon} is devoted to the basic characteristics of spontaneous emission of the multilevel atom.
In Sec.~\ref{sec:numerical} we present numerical results. 
Our conclusions are given in Sec.~\ref{sec:summary}.

\label{sec:introduction}

\section{Model and Hamiltonian}
\label{sec:model}

\begin{figure}[tbh]
\begin{center}
  \includegraphics{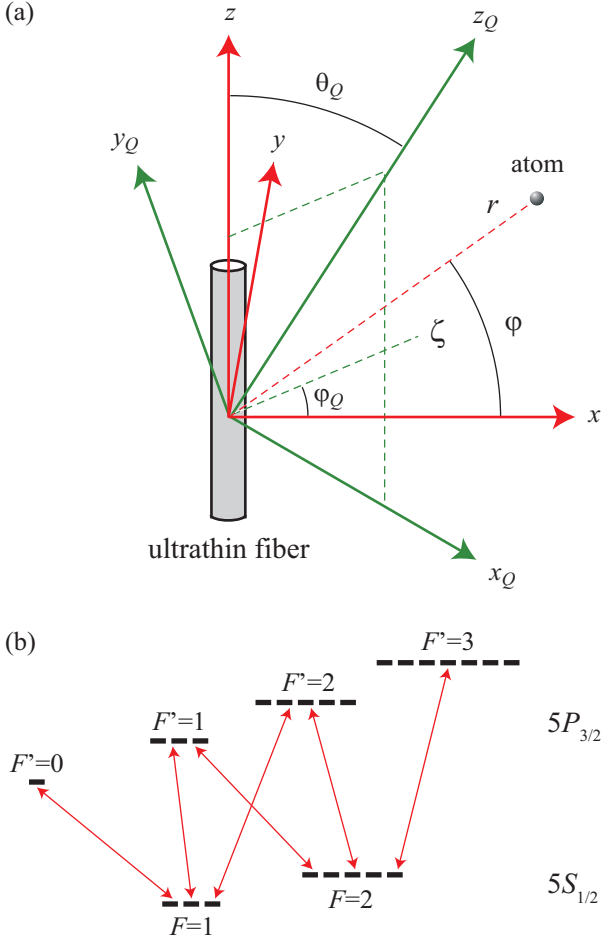}
 \end{center}
\caption{(Color online) (a) An atom interacting with guided and radiation modes of an ultrathin optical fiber.
The fiber-based Cartesian coordinate system $\{x,y,z\}$, the corresponding cylindrical coordinate system $\{r,\varphi,z\}$,
and the quantization coordinate system $\{x_Q,y_Q,z_Q\}$ are used.
(b) Schematic of the hfs levels of the $5P_{3/2}$  and $5S_{1/2}$ states of a rubidium-87 atom. These levels are specified  with respect to the quantization axis $z_Q$.}
\label{fig1}
\end{figure}

We consider a multilevel alkali-metal atom trapped in the vicinity of a vacuum-clad ultrathin optical fiber [see Fig.~\ref{fig1}(a)].
We use Cartesian coordinates $\{x,y,z\}$, where $z$ is the coordinate along the fiber axis, and also cylindrical coordinates $\{r,\varphi,z\}$, where $r$ and $\varphi$ are the polar coordinates in the fiber transverse plane $xy$. The energy levels of the atom are specified in a Cartesian coordinate system $\{x_Q,y_Q,z_Q\}$, where $z_Q$ is the direction of the quantization axis.

To be concrete, we assume that the atom is  $^{87}$Rb. 
We work with the $D_2$ line of the rubidium atom, which corresponds to the electric dipole transition from the excited state $5P_{3/2}$ to the ground state $5S_{1/2}$ [see Fig.~\ref{fig1}(b)] \cite{coolingbook}.
We introduce the notations $|e\rangle=|J'F'M'\rangle$ and $|g\rangle=|JFM\rangle$ for the magnetic sublevels of the hyperfine-structure (hfs) levels of the excited state and the ground state, 
respectively. Here, $J$ and $J'$ are the total electronic angular momenta,
$F$ and $F'$ are the total atomic angular momenta, and $M$ and $M'$ are the magnetic quantum numbers. 
We denote the energies of these sublevels as $\hbar\omega_{e}$ and $\hbar\omega_{g}$. 
The schematic of the hfs levels of the $D_2$ line of the rubidium-87 atom is illustrated in  Fig.~\ref{fig1}(b). 

We introduce the notation $\mathbf{d}_{eg}=\langle e|\mathbf{D}|g\rangle$ for the dipole
matrix element of the transition $|e\rangle\leftrightarrow|g\rangle$, where $\mathbf{D}$ is the electric dipole operator.
In the atomic quantization coordinate system $\{x_Q,y_Q,z_Q\}$, 
the spherical components  $q=0,\pm1$ of the  dipole matrix element $\mathbf{d}_{eg}$ are given by the expression \cite{Shore}
\begin{eqnarray}\label{1}
d_{q_Q}&=&(-1)^{I+J'-M'}\langle J' \| D\| J\rangle\sqrt{(2F+1)(2F'+1)}\nonumber\\
&&\mbox{}\times
\begin{Bmatrix}J'&F'&I\\F&J&1\end{Bmatrix}
\begin{pmatrix}F&1&F'\\M&q&-M'\end{pmatrix}.
\end{eqnarray}
Here, the array in the curly braces is a 6$j$ symbol, the array in the parentheses is a 3$j$ symbol, $I$ is the nuclear spin, 
and $\langle J' \| D\| J\rangle$ is the reduced electric dipole matrix element in the $J$ basis.
Note that $d_{q_Q}$ is nonzero only for $M'-M=q=0,\pm1$. 

We assume that the fiber has a cylindrical silica core of radius $a$ and refractive index $n_1$  
and  an infinite vacuum cladding of refractive index $n_2=1$. 
We retain the silica dispersion and at the frequency of the rubidium $D_2$ line the refractive index $n_1$ of the fiber  is taken as 1.4537.
The positive-frequency part $\mathbf{E}^{(+)}$ of the electric component of the field can be
decomposed into the contributions $\mathbf{E}^{(+)}_{\mathrm{g}}$ and $\mathbf{E}^{(+)}_{\mathrm{r}}$ from guided and radiation modes, respectively,  as  
\begin{equation}
\mathbf{E}^{(+)}=\mathbf{E}^{(+)}_{\mathrm{g}}+\mathbf{E}^{(+)}_{\mathrm{r}}.
\label{2}
\end{equation}
In view of the very low losses of silica in the wavelength range of interest, we neglect material absorption. 

We follow the continuum field quantization procedures presented in  \cite{Loudon}.
Regarding the guided modes, we assume that the fiber supports the fundamental HE$_{11}$ mode and a few higher-order modes \cite{fiber books} 
in a finite bandwidth around the central frequency $\omega_0=\omega_e-\omega_g$ of the rubidium-87 $D_2$ line. 
We label each guided mode in this bandwidth by an index $\mu=(\omega,N,f,p)$. 
Here, $\omega$ is the mode frequency, the notation $N=\mathrm{HE}_{lm}$, EH$_{lm}$, TE$_{0m}$, or TM$_{0m}$ stands for the mode type, with $l=1,2,\dots$ and $m=1,2,\dots$ being the azimuthal and radial mode orders, respectively, and the index $f=+1$ or $-1$ denotes respectively the forward or backward propagation direction along the fiber axis $z$. The HE$_{lm}$ and EH$_{lm}$ modes are hybrid modes. For these modes, the azimuthal order is $l\not=0$,
and the index $p$ is equal to $+1$ or $-1$, indicating the counterclockwise or clockwise circulation direction of the helical phasefront. The TE$_{0m}$ and TM$_{0m}$ modes are transverse electric and magnetic modes. For these modes, the azimuthal mode order is $l=0$ and, hence, the mode polarization is single
and the polarization index $p$ can take an arbitrary value. For convenience, we assign the value $p=0$ to the polarization index $p$ for TE$_{0m}$ and TM$_{0m}$ modes. In the interaction picture, the quantum expression for the positive-frequency part $\mathbf{E}^{(+)}_{\mathrm{g}}$ of the electric component of the field in guided modes is \cite{cesium decay}
\begin{equation}
\mathbf{E}^{(+)}_{\mathrm{g}}=i\sum_{\mu}\sqrt{\frac{\hbar\omega\beta'}{4\pi\epsilon_0}}
\;a_{\mu}\mathbf{e}^{(\mu)}e^{-i(\omega t-f\beta z-pl\varphi)}.
\label{3}
\end{equation}
Here, $\mathbf{e}^{(\mu)}=\mathbf{e}^{(\mu)}(r,\varphi)$ is the profile function of the guided mode $\mu$ in the classical problem, $a_{\mu}$ is the corresponding photon annihilation  operator, 
$\sum_{\mu}=\sum_{N fp}\int_0^{\infty}d\omega$ is the generalized summation over the guided modes,
$\beta$ is the longitudinal propagation constant, and $\beta'$ is the derivative of $\beta$
with respect to $\omega$. The constant $\beta$ is determined by the
fiber eigenvalue equation \cite{fiber books}. The operators $a_{\mu}$ and $a_{\mu}^\dagger$ satisfy the continuous-mode bosonic commutation rules $[a_{\mu},a_{\mu'}^\dagger]=\delta(\omega-\omega')\delta_{NN'}\delta_{ff'}\delta_{pp'}$. 
In deriving Eq.~\eqref{3}, we have used the normalization condition 
\begin{equation}\label{4}
\int _{0}^{2\pi}d\varphi\int _{0}^{\infty}n_{\mathrm{ref}}^2\,|\mathbf{e}^{(\mu)}|^2r\,dr=1,
\end{equation}
where $n_{\mathrm{ref}}(r)=n_1$ for $r<a$ and $n_2$ for $r>a$.

The explicit expressions for the profile functions $\mathbf{e}^{(\mu)}$ of guided modes are given in Refs.~\cite{fiber books,Fam2017} and are summarized in Appendix \ref{sec:guided}.
For a hybrid mode $N=\mathrm{HE}_{lm}$ and EH$_{lm}$ with the propagation direction $f$ and the phase circulation direction $p$, 
the profile function is given in the cylindrical coordinates as
\begin{equation}\label{5}
\mathbf{e}^{(\omega Nfp)}\big|_{N=\mathrm{HE}_{lm},\mathrm{EH}_{lm}}=e_r\hat{\mathbf{r}}+pe_\varphi\hat{\boldsymbol{\varphi}}+fe_z\hat{\mathbf{z}},
\end{equation}
where $e_r$, $e_\varphi$, and $e_z$ 
are given by Eqs.~\eqref{a10} and \eqref{a11} for $\beta>0$ and $l>0$. 
For a TE$_{0m}$ mode with the propagation direction $f$, the profile function can be written as
\begin{equation}\label{6}
\mathbf{e}^{(\omega \mathrm{TE}_{0m}f)}=\mathbf{e}^{(\omega \mathrm{TE}_{0m}fp)}\big|_{p=0}=e_\varphi\hat{\boldsymbol{\varphi}},
\end{equation}
where the only nonzero cylindrical component $e_\varphi$ is given by the second expressions in Eqs.~\eqref{a17} and \eqref{a18}.
For a TM mode with the propagation direction $f$, we have
\begin{equation}\label{7}
\mathbf{e}^{(\omega \mathrm{TM}_{0m}f)}=\mathbf{e}^{(\omega \mathrm{TM}_{0m}fp)}\big|_{p=0}=e_r\hat{\mathbf{r}}+fe_z\hat{\mathbf{z}},
\end{equation}
where the components $e_r$ and $e_z$ are given by the first and third expressions in Eqs.~\eqref{a22} and \eqref{a23} for $\beta>0$.
An important property of the mode functions of hybrid and TM modes is that the longitudinal
component $e_z$ is nonvanishing and in quadrature ($\pi/2$ out of phase) with the radial component $e_r$.

In the case of radiation modes, the longitudinal propagation constant $\beta$ for each value of the frequency $\omega$ can vary continuously, from $-k$ to $k$ (with $k=\omega/c$). We label each radiation mode by an index $\nu=(\omega,\beta,l,p)$, where 
$l=0,\pm1,\pm2,\dots$ is the mode order and  $p=+,-$ is the mode polarization. In the interaction picture, the quantum expression for the positive-frequency part $\mathbf{E}^{(+)}_{\mathrm{r}}$ of the electric component of the field in radiation modes is \cite{cesium decay}
\begin{equation}
\mathbf{E}^{(+)}_{\mathrm{r}}=i\sum_{\nu}
\sqrt{\frac{\hbar\omega}{4\pi\epsilon_0}}\;a_{\nu}\mathbf{e}^{(\nu)}e^{-i(\omega t-\beta z-l\varphi)}.
\label{8}
\end{equation}
Here, $\mathbf{e}^{(\nu)}=\mathbf{e}^{(\nu)}(r,\varphi)$ is the profile function of the radiation mode $\nu$ in the classical problem, $a_{\nu}$ is the corresponding photon annihilation 
operator, and $\sum_{\nu}=\sum_{lp}\int_0^{\infty}d\omega\int_{-k}^{k}d\beta$ is the generalized summation over the radiation modes. The operators $a_{\nu}$ and $a_{\nu}^\dagger$ satisfy the continuous-mode bosonic commutation rules $[a_{\nu},a_{\nu'}^\dagger]=\delta(\omega-\omega')\delta(\beta-\beta')
\delta_{ll'}\delta_{pp'}$. 
In deriving Eq.~\eqref{8}, we have used the normalization condition 
\begin{eqnarray}\label{9}
&&\int _0^{2\pi}d\varphi\int _{0}^{\infty}n_{\mathrm{ref}}^2
\left[\mathbf{e}^{(\nu)}\mathbf{e}^{(\nu')*}\right]_{\beta=\beta',l=l',p=p'}
r\,dr\nonumber\\&&
=\delta(\omega-\omega').
\end{eqnarray}
The explicit expressions for the mode functions $\mathbf{e}^{(\nu)}$ are given
in Refs. \cite{fiber books,Fam2017} and are summarized in Appendix \ref{sec:radiation}.

Assume that the atom is positioned at a point $(r,\varphi,z)$. The Hamiltonian for the atom-field interaction  in the dipole and rotating-wave approximations is given by 
\begin{equation}\label{10}
H_{\mathrm{int}}=-i\hbar\sum_{\alpha eg}G_{\alpha eg}
\sigma_{ge}^\dagger a_{\alpha}e^{-i(\omega-\omega_{eg})t}+\mbox{H.c.},
\end{equation}
where the notations $\alpha=\mu,\nu$ and $\sum_{\alpha}=\sum_{\mu}+\sum_{\nu}$   
stand for the general mode index and the complete mode summation, respectively, and
the operators $\sigma_{ge}=|g\rangle\langle e|$ and 
$\sigma_{ge}^\dagger=\sigma_{eg}=|e\rangle\langle g|$
describe the downward and upward transitions, respectively. 
The coefficients 
\begin{eqnarray}\label{11}
G_{\mu eg}&=&\sqrt{\frac{\omega\beta'}{4\pi\epsilon_0\hbar}}\;
\big(\mathbf{d}_{eg}\cdot\mathbf{e}^{(\mu)}\big)e^{i(f\beta z+pl\varphi)},\nonumber\\
G_{\nu eg}&=&\sqrt{\frac{\omega}{4\pi\epsilon_0\hbar}}\;
\big(\mathbf{d}_{eg}\cdot\mathbf{e}^{(\nu)}\big)e^{i(\beta z+l\varphi)}
\end{eqnarray}
characterize the coupling of the atomic transition $e\leftrightarrow g$ with
the guided mode $\mu$ and the radiation mode $\nu$.   
The notation $\omega_{eg}=\omega_e-\omega_g$ stands for the atomic transition frequency. 

We note that, for $|e\rangle=|J'F'M'\rangle$ and $|g\rangle=|JFM\rangle$, 
the scalar product of the atomic dipole vector $\mathbf{d}_{eg}$ and the field vector $\mathbf{e}^{(\alpha)}$ can be expressed as
$\mathbf{d}_{eg}\cdot\mathbf{e}^{(\alpha)}=(-1)^{q} d_{q_Q}e_{-q_Q}^{(\alpha)}\big|_{q=M'-M}$, where $d_{q_Q}$ is given by Eq. (\ref{1}) 
and $e_{q_Q}^{(\alpha)}$ is the corresponding spherical tensor component of the field in the atomic quantization coordinate system $\{x_Q,y_Q,z_Q\}$.
The components $e_{q_Q}^{(\alpha)}$ with $q=0,\pm1$ are defined as $e_{-1_Q}^{(\alpha)}=(e_{x_Q}^{(\alpha)}-ie_{y_Q}^{(\alpha)})/\sqrt{2}$, $e_{0_Q}^{(\alpha)}=e_{z_Q}^{(\alpha)}$, and $e_{1_Q}^{(\alpha)}=-(e_{x_Q}^{(\alpha)}+ie_{y_Q}^{(\alpha)})/\sqrt{2}$. 
Let $\theta_Q$ be the angle between the quantization axis $z_Q$ and the fiber axis $z$  [see Fig.~\ref{fig1}(a)].
Assume that the plane $(z,z_Q)$ intersects with the fiber transverse plane $xy$ at a line $\zeta$. Let $\varphi_Q$ be the azimuthal angle between $\zeta$ and $x$. We choose the axes $x_Q$ and $y_Q$ such that $x_Q$ is in the plane $(z_Q,z)$ and $y_Q$ is in the plane $(x,y)$. 
Then, the transformation for the field vector $\mathbf{e}^{(\alpha)}$ from the coordinate system $\{x,y,z\}$ to the coordinate system $\{x_Q,y_Q,z_Q\}$ is given by the equations
\begin{eqnarray}\label{12}
e_{x_Q}^{(\alpha)}&=&(e_x^{(\alpha)}\cos\varphi_Q+e_y^{(\alpha)}\sin\varphi_Q)\cos\theta_Q-e_z^{(\alpha)}\sin\theta_Q,\nonumber\\ 
e_{y_Q}^{(\alpha)}&=&-e_x^{(\alpha)}\sin\varphi_Q+e_y^{(\alpha)}\cos\varphi_Q,\nonumber\\  
e_{z_Q}^{(\alpha)}&=&(e_x^{(\alpha)}\cos\varphi_Q+e_y^{(\alpha)}\sin\varphi_Q)\sin\theta_Q+e_z^{(\alpha)}\cos\theta_Q.\nonumber\\
\end{eqnarray}
The relations between the Cartesian-coordinate vector components $e_x^{(\alpha)}$ and $e_y^{(\alpha)}$ and the cylindrical-coordinate vector components $e_r^{(\alpha)}$ and $e_\varphi^{(\alpha)}$ are $e_x^{(\alpha)}=e_r^{(\alpha)}\cos\varphi-e_\varphi^{(\alpha)}\sin\varphi$ and $e_y^{(\alpha)}=e_r^{(\alpha)}\sin\varphi+e_\varphi^{(\alpha)}\cos\varphi$.

\section{Spontaneous emission of the atom}
\label{sec:spon}

In this section, we study spontaneous emission of the multilevel atom.
We assume that the field is initially in the vacuum state $|0\rangle$. In this case,
the time evolution of the reduced density operator $\rho$ of the atom
is governed by the equations \cite{cesium decay}
\begin{eqnarray}\label{13}
\dot{\rho}_{ee'}&=&
-\frac{1}{2}\sum_{e''}(\gamma_{ee''}{\rho}_{e''e'}+\gamma_{e''e'}{\rho}_{ee''}),\nonumber\\
\dot{\rho}_{gg'}&=&
\sum_{ee'}\gamma_{e'eg'g}{\rho}_{ee'},\nonumber\\
\dot{\rho}_{eg}&=& 
-\frac{1}{2}\sum_{e'} \gamma_{ee'}\rho_{e'g},
\end{eqnarray}
where the coefficients 
\begin{equation}\label{14}
\begin{split}
\gamma_{ee'gg'}&=\gamma^{(\mathrm{g})}_{ee'gg'}+\gamma^{(\mathrm{r})}_{ee'gg'},\\
\gamma_{ee'}&=\sum_{g}\gamma_{ee'gg}=\gamma^{(\mathrm{g})}_{ee'}+\gamma^{(\mathrm{r})}_{ee'}
\end{split}
\end{equation}
characterize the spontaneous emission process.
In Eqs.~\eqref{14}, the set of coefficients $\gamma^{(\mathrm{g})}_{ee'gg'}$ and $\gamma^{(\mathrm{g})}_{ee'}$ describes spontaneous emission into guided modes, and the set of coefficients $\gamma^{(\mathrm{r})}_{ee'gg'}$ and 
$\gamma^{(\mathrm{r})}_{ee'}$ describes spontaneous emission into radiation modes. 
The expressions for these coefficients are given as \cite{cesium decay} 
\begin{eqnarray}\label{15}
\gamma^{(\mathrm{g})}_{ee'gg'}&=&2\pi \sum_{Nfp}G_{Nfp eg}G_{Nfp e'g'}^*, \nonumber\\
\gamma^{(\mathrm{g})}_{ee'}&=&\sum_{g}\gamma^{(\mathrm{g})}_{ee'gg},
\end{eqnarray}
and
\begin{eqnarray}\label{16}
\gamma^{(\mathrm{r})}_{ee'gg'}&=&2\pi \sum_{lp}\int_{-k_0n_2}^{k_0n_2}d\beta\,G_{\beta lp eg}G_{\beta lp e'g'}^*,
\nonumber\\
\gamma^{(\mathrm{r})}_{ee'}&=&\sum_{g}\gamma^{(\mathrm{r})}_{ee'gg},
\end{eqnarray}
where $G_{Nfpeg}\equiv G_{\omega_0Nfpeg}$ and 
$G_{\beta lpeg}\equiv G_{\omega_0\beta lpeg}$ are the coupling coefficients for the resonant guided and radiation modes, respectively.

The diagonal decay coefficients $\gamma^{(\mathrm{g})}_{e}\equiv\gamma^{(\mathrm{g})}_{ee}$ and $\gamma^{(\mathrm{r})}_{e}=\gamma^{(\mathrm{r})}_{ee}$ are the rates of spontaneous emission from the magnetic sublevel $|e\rangle$ of the atom into guided and radiation modes, respectively.
The total decay rate for the population of the sublevel $|e\rangle$ is
\begin{equation}\label{17} 
\gamma_{e}\equiv\gamma_{ee}=\gamma^{(\mathrm{g})}_{e}+\gamma^{(\mathrm{r})}_{e}.
\end{equation}
The rate of spontaneous emission from the magnetic sublevel $|e\rangle$ of the atom into the guided modes $N=$ HE$_{lm}$, EH$_{lm}$, TE$_{0m}$, or TM$_{0m}$ is given by
\begin{equation}\label{18} 
\gamma^{(N)}_{e}=2\pi \sum_{fpg}|G_{Nfp eg}|^2.
\end{equation}
It is clear that 
\begin{equation}\label{19} 
\gamma^{(\mathrm{g})}_{e}=\sum_N\gamma^{(N)}_{e}=2\pi \sum_{Nfpg}|G_{Nfp eg}|^2.
\end{equation}

We note that the density-matrix equations \eqref{13} are in agreement with those used in the treatments for the excitation of a multilevel atom by light of arbitrary polarization \cite{Milner98,Milner99,Taichenachev99,Vitanov03,Taichenachev04,Yudin13,ChangMinogin}. Equations \eqref{13} can, in principle, be used for an arbitrary (degenerate and nondegenerate) multilevel atom. The tensor nature of the Zeeman sublevels and the hfs levels of a realistic alkali-metal atom is expressed by Eq. \eqref{1} for the spherical tensor components of the atomic dipole matrix elements $\mathbf{d}_{eg}$. 
These quantities enter Eqs. \eqref{13} through expressions \eqref{11} for the coupling coefficients $G_{\mu eg}$ and $G_{\nu eg}$.
Unlike the case of the atom-field system in free space \cite{ChangMinogin}, the presence of the nanofiber modifies the decay rates $\gamma_{e}$ and leads to the appearance of the cross-level decay coefficients $\gamma_{ee'}$ (with $e\not= e'$) in Eqs.~\eqref{13} (see \cite{cesium decay}).

We introduce the notation
\begin{equation}\label{20} 
\gamma^{(Nfp)}_{eg}=2\pi|G_{Nfp eg}|^2,
\end{equation}
which stands for the rate of spontaneous emission into the guided modes $Nfp$ via the transition $|e\rangle\to|g\rangle$.
The rate of spontaneous emission from the sublevel $|e\rangle$ of the atom into the guided modes $N$ with the propagation direction 
$f$ is given by
\begin{equation}\label{21} 
\gamma^{(Nf)}_{e}=\sum_{pg}\gamma^{(Nfp)}_{eg}.
\end{equation}
The rate of spontaneous emission into all types of guided modes propagating in the direction $f$ is given by
\begin{equation}\label{22} 
\gamma^{(\mathrm{g}f)}_{e}=\sum_N\gamma^{(Nf)}_{e}.
\end{equation}

For TE modes, the profile function for the electric part of the field does not depend on the propagation direction $f$ [see Eq.~\eqref{6}].
Therefore, the rates $\gamma^{(Nf)}_{e}$ for $N=\mathrm{TE}$ modes is symmetric with respect to $f$.
 
For hybrid and TM modes, the longitudinal component $e_z^{(\omega Nfp)}$ of the field is nonvanishing and has opposite signs for opposite propagation directions [see Eqs.~\eqref{5} and \eqref{7}].
Therefore, the rates $\gamma^{(Nf)}_{e}$ for $N=$ HE, EH, and TM modes and the rate $\gamma^{(\mathrm{g}f)}_{e}$ for all guided modes may depend on $f$. 

The rates $\gamma^{(Nf)}_{e}$ and hence $\gamma^{(\mathrm{g}f)}_{e}$ do not depend on $f$ 
when the quantization axis $z_Q$ coincides with the fiber axis $z$. 
Indeed, in this case, we have $e_{q_Q}^{(\omega Nfp)}=e_q^{(\omega Nfp)}$ for $q=0,\pm1$, where $e_q^{(\omega Nfp)}$ are
the spherical tensor components of the mode function $\mathbf{e}^{(\omega Nfp)}$ in the fiber coordinate system $\{x,y,z\}$. 
These components satisfy the relation 
\begin{equation}\label{23} 
e_q^{(\omega Nfp)}=(-1)^{1+q}e_q^{(\omega N\bar{f}p)},
\end{equation}
where $\bar{f}=-f$. Hence, we find the relation
\begin{equation}\label{24} 
G_{Nfpeg}=(-1)^{1+M'-M}G_{N\bar{f}peg}e^{2if\beta z},
\end{equation}
which yields $\gamma^{(Nfp)}_{eg}=\gamma^{(N\bar{f}p)}_{eg}$ and, hence, $\gamma^{(N+)}_{e}=\gamma^{(N-)}_{e}$ and $\gamma^{(\mathrm{g}+)}_{e}=\gamma^{(\mathrm{g}-)}_{e}$.

More generally, we find that $\gamma^{(Nf)}_{e}$ and hence $\gamma^{(\mathrm{g}f)}_{e}$ do not depend on $f$ 
when the quantization axis $z_Q$ lies in the meridional plane containing the position of the atom.
In order to show this directional independence, we assume that the atom is on the $x$ axis and the quantization axis $z_Q$ lies in the $zx$ plane, that is, $\varphi_Q=0$. 
Then, for hybrid and TM modes with the profile functions \eqref{5} and \eqref{7}, Eqs.~\eqref{12} yield
\begin{eqnarray}\label{25}
e_{x_Q}^{(\mu)}&=&e_r\cos\theta_Q-fe_z\sin\theta_Q,\nonumber\\ 
e_{y_Q}^{(\mu)}&=&pe_\varphi,\nonumber\\  
e_{z_Q}^{(\mu)}&=&e_r\sin\theta_Q+fe_z\cos\theta_Q.
\end{eqnarray}
According to Appendix \ref{sec:guided}, for an appropriate choice of the normalization constant, 
$e_z$ and $e_\varphi$ are real numbers and $e_r$ is an imaginary number. Hence, we can show that
the absolute values $|e_{q_Q}^{(\mu)}|$ of the spherical tensor components of the field in the
coordinate system $\{x_Q,y_Q,z_Q\}$ do not depend on $f$. On the other hand, the dipole matrix element $\mathbf{d}_{eg}$
has a single nonzero spherical tensor component $d_{q_Q}$, which is a real number. Consequently, the absolute value of the scalar product 
$\mathbf{d}_{eg}\cdot\mathbf{e}^{(\mu)}$ is $|\mathbf{d}_{eg}\cdot\mathbf{e}^{(\mu)}|=|d_{q_Q}||e_{-q_Q}^{(\mu)}|$. 
This quantity is independent of $f$ and, hence, so are the rates $\gamma^{(Nf)}_{e}$ and $\gamma^{(\mathrm{g}f)}_{e}$.

It is worth noting that the rates $\gamma^{(N)}_{e}$, $\gamma^{(\mathrm{g})}_{e}$, $\gamma^{(\mathrm{r})}_{e}$, and 
$\gamma_{e}$ are symmetric with respect to the magnetic quantum number $M'$ of the sublevel $|e\rangle=|F'M'\rangle=|J'F'M'\rangle$, that is, $\gamma^{(N)}_{e}=\gamma^{(N)}_{\bar{e}}$, $\gamma^{(\mathrm{g})}_{e}=\gamma^{(\mathrm{g})}_{\bar{e}}$, $\gamma^{(\mathrm{r})}_{e}=\gamma^{(\mathrm{r})}_{\bar{e}}$,
and $\gamma_{e}=\gamma_{\bar{e}}$, where the index $\bar{e}$ labels the sublevel $|\bar{e}\rangle=|F',-M'\rangle$ with the opposite magnetic quantum number $-M'$.
This symmetry is a consequence of the properties
\begin{equation}\label{26}
\begin{split} 
\mathbf{e}^{(\omega, N, f,p)}&=-\mathbf{e}^{(\omega, N,-f,-p)*},\\
\mathbf{e}^{(\omega,\beta,l,p)}&=(-1)^l\mathbf{e}^{(\omega,-\beta,-l,p)*},
\end{split}
\end{equation} 
and 
\begin{equation}\label{27}
\mathbf{d}_{eg}=(-1)^{F'-F+M'-M+1}\mathbf{d}_{\bar{e}\bar{g}}^*, 
\end{equation} 
where $|\bar{g}\rangle=|F,-M\rangle$.
With the help of the relations \eqref{26} and \eqref{27}, we can also show that
\begin{equation}\label{28}
\begin{split}
\gamma^{(Nf)}_{e}&=\gamma^{(N\bar{f})}_{\bar{e}},\\
\gamma^{(\mathrm{g}f)}_{e}&=\gamma^{(\mathrm{g}\bar{f})}_{\bar{e}}.
\end{split}
\end{equation}
Thus, the rates $\gamma^{(Nf)}_{e}$ and $\gamma^{(\mathrm{g}f)}_{e}$ of spontaneous emission into guided modes propagating in a given direction $f$ do not change when both the propagation direction $f$ and the magnetic quantum number $M'$ are reversed.
It is clear that if $\gamma^{(Nf)}_{e}$ and $\gamma^{(\mathrm{g}f)}_{e}$ depend on $f$ then they also depend on the sign of $M'$ and vice versa.  

In order to get insight into the direction dependencies of the spontaneous emission rates, we consider  
the rate $\gamma^{(Nf)}_{eg}\equiv \sum_p\gamma^{(Nfp)}_{eg}$ for a given transition $|e\rangle\to|g\rangle$. 
When we follow the procedure of Ref.~\cite{flat}, we can decompose this rate as
\begin{equation}\label{fl29}
\gamma^{(Nf)}_{eg}=\gamma_0^{(f)}+\gamma_1^{(f)}+\gamma_2^{(f)},
\end{equation}
where
\begin{equation}\label{fl30}
\begin{split}
\gamma_0^{(f)}&=\frac{\omega_0\beta_0'}{6\epsilon_0\hbar}|\mathbf{d}_{eg}|^2\sum_{p}|\mathbf{e}^{(\omega_0Nfp)}|^2,\\
\gamma_1^{(f)}&=\frac{\omega_0\beta_0'}{4\epsilon_0\hbar}[\mathbf{d}_{eg}^*\times\mathbf{d}_{eg}]\cdot\sum_{p}[\mathbf{e}^{(\omega_0Nfp)*}\times\mathbf{e}^{(\omega_0Nfp)}],\\
\gamma_2^{(f)}&=\frac{\omega_0\beta_0'}{2\epsilon_0\hbar}\{\mathbf{d}_{eg}^*\otimes\mathbf{d}_{eg}\}_2\cdot\sum_{p}\{\mathbf{e}^{(\omega_0Nfp)*}\otimes\mathbf{e}^{(\omega_0Nfp)}\}_2.
\end{split}
\end{equation}
Here, the notation $\{\mathbf{A}\otimes\mathbf{B}\}_{2}$ stands for the irreducible tensor product of rank 2 of arbitrary complex vectors $\mathbf{A}$ and $\mathbf{B}$. The quantities $\gamma_0^{(f)}$, $\gamma_1^{(f)}$, and  $\gamma_2^{(f)}$ are called the scalar, vector, and tensor components of the rate $\gamma^{(Nf)}_{eg}$, respectively. 

With the help of the first relation in Eqs.~\eqref{26}, we can show that $\gamma_0^{(f)}=\gamma_0^{(\bar{f})}$, $\gamma_1^{(f)}=-\gamma_1^{(\bar{f})}$, and $\gamma_2^{(f)}=\gamma_2^{(\bar{f})}$. Thus, the direction dependence of the rate $\gamma^{(Nf)}_{eg}$
occurs when the vector term $\gamma_1^{(f)}$ is nonvanishing.

According to the second expression in  Eqs.~\eqref{fl30}, the vector term $\gamma_1^{(f)}$ depends on the overlap between the vectors $i[\mathbf{d}_{eg}^*\times\mathbf{d}_{eg}]$ and $-i[\mathbf{e}^{(\omega_0Nfp)*}\times\mathbf{e}^{(\omega_0Nfp)}]$, which are proportional to the ellipticity vector of the atomic electric dipole polarization and the ellipticity vector of the electric field polarization, respectively. 
The vector $i[\mathbf{d}_{eg}^*\times\mathbf{d}_{eg}]$ characterizes an effective magnetic dipole produced by the rotation of the electric dipole, and is responsible for the vector polarizability of the atom.
The vector $-i[\mathbf{e}^{(\omega_0Nfp)*}\times\mathbf{e}^{(\omega_0Nfp)}]$ characterizes an effective magnetic field and is responsible for the local electric spin  density of light. The vector component $\gamma_1^{(f)}$ of the rate can be considered as a result of the interaction between the effective magnetic dipole and the effective magnetic field. Due to spin-orbit coupling of light \cite{Zeldovich,Bliokh review,Bliokh review2015,Bliokh2014,Bliokh2015}, a reverse of the propagation direction leads to a reverse of the spin density of light and, consequently, to a reverse of the vector component $\gamma_1^{(f)}$ of the spontaneous emission rate $\gamma^{(Nf)}_{eg}$.

We can show that $\sum_p[\mathbf{e}^{(\omega_0Nfp)*}\times\mathbf{e}^{(\omega_0Nfp)}]\propto fe_ze_r\hat{\boldsymbol{\varphi}}$,
which leads to $\gamma_1^{(f)}\propto fe_ze_r([\mathbf{d}_{eg}^*\times\mathbf{d}_{eg}]\cdot\hat{\boldsymbol{\varphi}})$. 
Hence, the spontaneous emission rate $\gamma^{(Nf)}_{eg}$ depends on $f$ only when the ellipticity vector $i[\mathbf{d}_{eg}^*\times\mathbf{d}_{eg}]$ of the atomic dipole has a nonvanishing azimuthal component $i[\mathbf{d}_{eg}^*\times\mathbf{d}_{eg}]_\varphi$. 
It is clear that the direction dependence of $\gamma^{(Nf)}_{eg}$ is a consequence of the fact that the longitudinal component $e_z$ of the guided field is not zero.

\section{Numerical results}
\label{sec:numerical}

In this section, we demonstrate the results of numerical calculations for the  
decay characteristics of the magnetic sublevels of the excited state $5P_{3/2}$ of a rubidium-87 atom in the presence of an ultrathin optical fiber.
The atomic transitions between this state and the ground state $5S_{1/2}$ correspond to the $D_2$ line and have a wavelength $\lambda_0=780$ nm. 
For simplicity, we show only the results of calculations for the spontaneous emission rates $\gamma_{e}$ of the sublevels 
$|e\rangle=|F'M'\rangle=|J'F'M'\rangle$ and their components.

\subsection{Dependencies of the rates on the radial distance}
\label{subsec:radial}

In this subsection, we study the dependencies of the rates on the radial distance $r$.
For simplicity, we consider the case where the fiber axis $z$ is used as the quantization axis.
In this case, none of the rates depend on the azimuthal angle $\varphi$. In addition, the  decay rates  of the sublevels with the magnetic quantum numbers $M'$ and $-M'$ are the same.

\begin{figure}[tbh]
\begin{center}
 \includegraphics{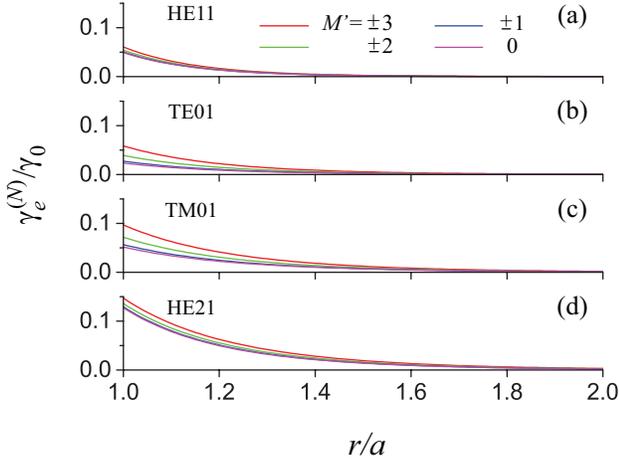}
 \end{center}
\caption{(Color online) Radial dependencies of the rates $\gamma_{e}^{(N)}$
of spontaneous emission from different magnetic sublevels of the hfs level $5P_{3/2}F'=3$ of a rubidium-87 atom into different guided modes of an ultrathin optical fiber. The quantization axis $z_Q$ coincides with the fiber axis $z$. 
The fiber radius is $a=400$ nm. The wavelength of the atomic transition is $\lambda_0=780$ nm. The refractive indices of the fiber and the vacuum cladding are $n_1=1.4537$ and $n_2=1$, respectively. The rates are normalized to the free-space decay rate $\gamma_0$ of the atom.}
\label{fig2}
\end{figure}

We show in Fig.~\ref{fig2} the radial dependencies of the rates $\gamma_{e}^{(N)}$ of spontaneous emission 
from different  magnetic sublevels of the hfs level $5P_{3/2}F'=3$ of the rubidium atom into different guided modes. The fiber radius is chosen to be $a=400$ nm. For the wavelength $\lambda_0=780$ nm, this fiber can support the HE$_{11}$, TE$_{01}$, TM$_{01}$, and HE$_{21}$ modes. 
According to Fig.~\ref{fig2}, the presence of the fiber leads to substantial decay rates into guided modes. Comparison between the different parts of the figure  shows that the emission into the HE$_{21}$ modes is stronger than into the HE$_{11}$, TE$_{01}$, and 
TM$_{01}$ modes. We observe that different magnetic sublevels have different decay rates, unlike the case of alkali-metal atoms in free space. 
The rates of spontaneous emission from the outermost magnetic sublevels $|F'=3,M'=\pm3\rangle$ (red lines) into guided modes are larger than those from the other sublevels. This indicates that the polarization profiles of the guided modes are more favorable to the $\sigma_{\pm}$ transitions than the $\pi$ transition.
The rates of spontaneous emission into guided modes are largest when the atom is positioned on the fiber surface. 
When the atom is far away from the fiber, $\gamma^{(N)}_{e}$ reduces to zero. Since the decay rates  of the sublevels $M'$ and $-M'$ are the same in the case where the quantization axis is the fiber axis, the maximum number of lines in each part of Fig.~\ref{fig2} is four. 
Since the difference between the  decay rates for $M'=0$ and $M'=\pm1$ is very small, we can clearly distinguish  only three lines in Figs.~\ref{fig2}(a) and \ref{fig2}(d). 

We note that our results presented in Fig.~\ref{fig2} do not agree quantitatively with the results of Masalov and Minogin \cite{Minogin2014}.
Indeed, the ratio between the rates of emission from the outermost levels into the HE$_{21}$ and HE$_{11}$ modes at the distance $r/a=1$ is equal to about 3 
in Fig.~\ref{fig2} but is equal to about 8 in the calculations of Ref.~\cite{Minogin2014}.
One of the reasons for the discrepancy is that they considered $^{85}$Rb, while we study $^{87}$Rb. Another reason is that they
limited their calculations to atomic transitions and guided modes with a single type of polarization, while we include all atomic transitions and field modes in our treatment. The most important reason for the discrepancy is that Eq.~(16) of Ref.~\cite{Minogin2014} is not accurate.     

\begin{figure}[tbh]
\begin{center}
 \includegraphics{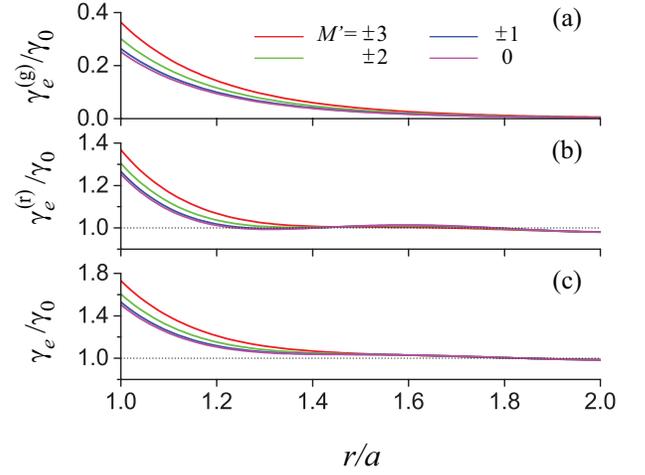}
 \end{center}
\caption{(Color online) Radial dependencies of the rates $\gamma_{e}^{(\mathrm{g})}$, $\gamma_{e}^{(\mathrm{r})}$, and $\gamma_{e}$
of spontaneous emission from different magnetic sublevels of the hfs level $5P_{3/2}F'=3$ into (a) guided modes, (b) radiation modes,
and (c) both types of modes. The parameters used are the same as for Fig.~\ref{fig2}.  The rates are normalized to the free-space decay rate $\gamma_0$ of the atom.
The dotted lines stand for unity and are guides to the eye.
}
\label{fig3}
\end{figure}

We show in Fig.~\ref{fig3} the radial dependencies of the spontaneous emission rates $\gamma_{e}^{(\mathrm{g})}$, $\gamma_{e}^{(\mathrm{r})}$, and $\gamma_{e}$ from different magnetic sublevels of the hfs level $5P_{3/2}F'=3$ into guided modes, radiation modes, and  both types of modes, respectively. 
We observe from Fig.~\ref{fig3}(a) that the rates $\gamma_{e}^{(\mathrm{g})}$ for the outermost sublevels $M'=\pm3$ (red lines) are larger than for the other sublevels. When the radial distance $r$ is not too large, the rates $\gamma_{e}^{(\mathrm{r})}$ and $\gamma_{e}$ for the sublevels $M'=\pm3$ are also larger than for the other sublevels [see Figs.~\ref{fig3}(b) and \ref{fig3}(c)]. When the atom is far away from the fiber, $\gamma^{(\mathrm{g})}_{e}$ reduces to zero [see Fig.~\ref{fig3}(a)], while $\gamma^{(\mathrm{r})}_{e}$ and $\gamma_{e}$ approach the free-space limiting value $\gamma_0$ [see Figs.~\ref{fig3}(b) and \ref{fig3}(c)]. The small oscillations around the value of unity in Fig.~\ref{fig3}(b) for $\gamma^{(\mathrm{r})}_{e}$ can be ascribed to the constructive and destructive interference due to reflections from the fiber surface \cite{Tromborg}. Due to the interference, the total rate $\gamma_{e}$ can become slightly smaller than $\gamma_0$ in some regions outside the fiber [see Fig.~\ref{fig3}(c)]. 

\begin{figure}[tbh]
\begin{center}
 \includegraphics{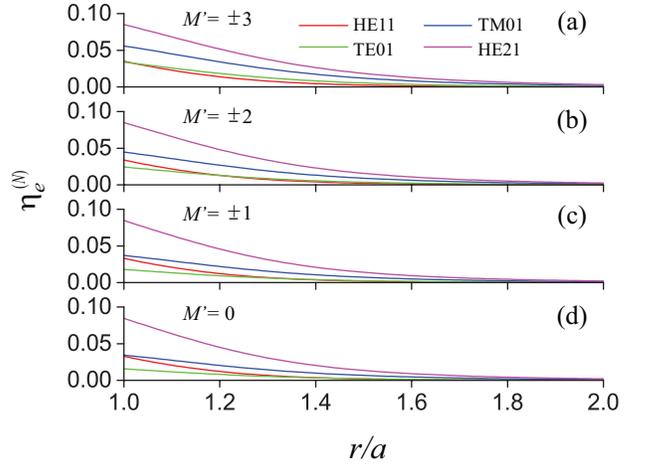}
 \end{center}
\caption{(Color online) Radial dependencies of the fractional rates $\eta_e^{(N)}=\gamma_{e}^{(N)}/\gamma_{e}$
of spontaneous emission from different magnetic sublevels of the hfs level $5P_{3/2}F'=3$ into different guided modes. 
The parameters used are the same as for Fig.~\ref{fig2}.
}
\label{fig4}
\end{figure}

We show in Fig.~\ref{fig4} the radial dependencies of the fractional rates $\eta_e^{(N)}=\gamma_{e}^{(N)}/\gamma_{e}$ of spontaneous emission from different magnetic sublevels of the hfs level $5P_{3/2}F'=3$ into different guided modes. The figure shows that 
the fractional rates $\eta_e^{(N)}$ of emission from the sublevels into the HE$_{21}$ modes are larger than those into the HE$_{11}$, TE$_{01}$, and TM$_{01}$ modes.

\begin{figure}[tbh]
\begin{center}
 \includegraphics{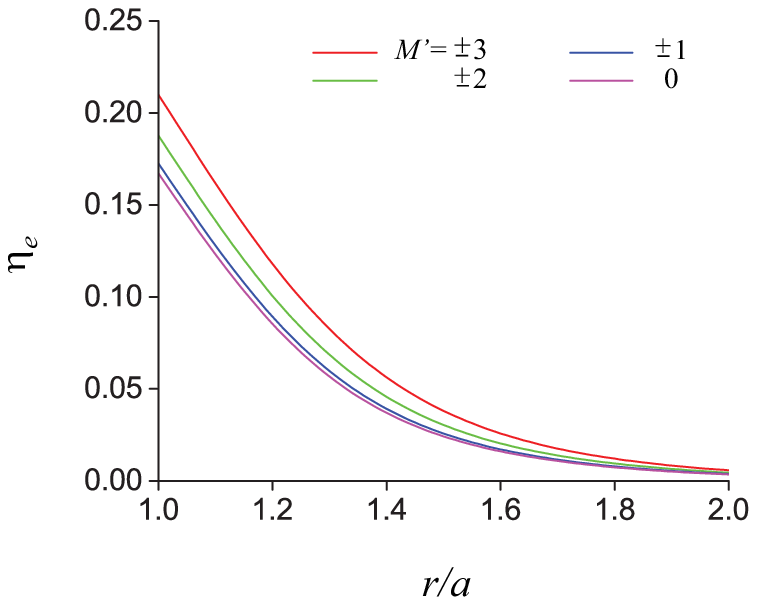}
 \end{center}
\caption{(Color online) Radial dependencies of the fractional rates $\eta_e=\gamma_{e}^{(\mathrm{g})}/\gamma_{e}$
of spontaneous emission from different magnetic sublevels of the hfs level $5P_{3/2}F'=3$ into all types of guided modes. 
The parameters used are the same as for Fig.~\ref{fig2}.
}
\label{fig5}
\end{figure}

We show in Fig.~\ref{fig5} the radial dependencies of the fractional rates $\eta_e=\gamma_{e}^{(\mathrm{g})}/\gamma_{e}=\sum_N\eta_e^{(N)}$ of spontaneous emission from different magnetic sublevels of the hfs level $5P_{3/2}F'=3$ into all types of guided modes. The figure shows that 
the outermost sublevels $M'=\pm3$ have the largest fractional rate. 
At the fiber surface, the fractional rates are largest. 
Their magnitudes are substantial, in the range from $0.17$ to $0.21$, depending on the magnetic quantum number $M'$.

\begin{figure}[tbh]
\begin{center}
 \includegraphics{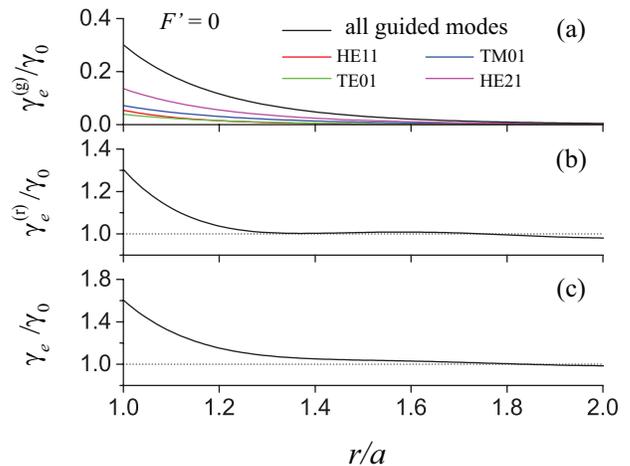}
 \end{center}
\caption{(Color online) Radial dependencies of the rates $\gamma_{e}^{(\mathrm{g})}$, $\gamma_{e}^{(\mathrm{r})}$, and $\gamma_{e}$
of spontaneous emission from the hfs level $5P_{3/2}F'=0$ into (a) guided modes, (b) radiation modes,
and (c) both types of modes. The components $\gamma_{e}^{(N)}$ of the rate $\gamma_{e}^{(\mathrm{g})}$ are also shown in part (a). 
The parameters used are the same as for Fig.~\ref{fig2}. The rates are normalized to the free-space decay rate $\gamma_0$ of the atom.
The dotted lines stand for unity and are guides to the eye.
}
\label{fig6}
\end{figure}

Note that the hfs level $5P_{3/2}F'=0$ is a singlet state, $|F'=0,M'=0\rangle$, which is equally coupled to the sublevels $|F=1,M=0,\pm1\rangle$ of the hfs level $F=1$ of the ground state $5S_{1/2}$. Therefore, the decay rate for the state $|F'=0,M'=0\rangle$ is equal to the average decay rate for an ensemble of two-level emitters with dipoles oriented randomly in space. We show in Fig.~\ref{fig6} the radial dependencies of the spontaneous emission rates $\gamma_{e}^{(\mathrm{g})}$, $\gamma_{e}^{(\mathrm{r})}$, and $\gamma_{e}$ from the hfs level $5P_{3/2}F'=0$ into guided modes, radiation modes, and  both types of modes.

\subsection{Dependencies of the rates on the fiber radius}
\label{subsec:radius}

In this subsection, we study the dependencies of the decay rates on the fiber radius $a$. We again use the fiber axis $z$ as the quantization axis.

\begin{figure}[tbh]
\begin{center}
 \includegraphics{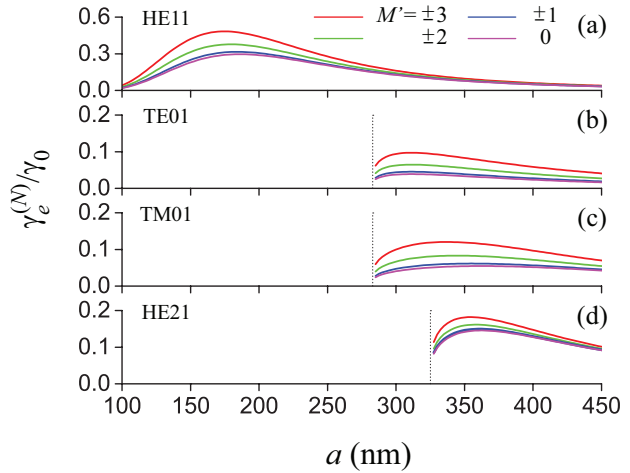}
 \end{center}
\caption{(Color online) 
Rates $\gamma^{(N)}_{e}$ of spontaneous emission from different magnetic sublevels of the hfs level $5P_{3/2}F'=3$ into different guided modes as functions of the fiber radius $a$. The atom is positioned on the fiber surface. Other parameters are as for Fig.~\ref{fig2}. The rates are normalized to the free-space decay rate $\gamma_0$ of the atom. The vertical dotted lines indicate the positions of the cutoffs for higher-order modes.
}
\label{fig7}
\end{figure}

In Fig.~\ref{fig7}, we show the rates $\gamma^{(N)}_{e}$ of spontaneous emission from different magnetic sublevels of the hfs level $5P_{3/2}F'=3$
into different guided modes as functions of the fiber radius $a$. 
We observe from the figure that the rates $\gamma^{(N)}_{e}$ have  maxima, whose positions and magnitudes strongly depend on the mode type $N$. 
The emission from the atom into the fundamental HE$_{11}$ modes is strongest when $a$ is around 180 nm.
For a given fiber radius $a$ in the range from 330 nm to 450 nm (the sizes that are typically achieved experimentally), the emission into the HE$_{21}$ modes is stronger than into the TM$_{01}$,
TE$_{01}$, and HE$_{11}$ modes. When the atom is positioned on the fiber surface, the rates $\gamma^{(N)}_{e}$ for the outermost sublevels $M'=\pm3$ are larger than for the other sublevels.

\begin{figure}[tbh]
\begin{center}
 \includegraphics{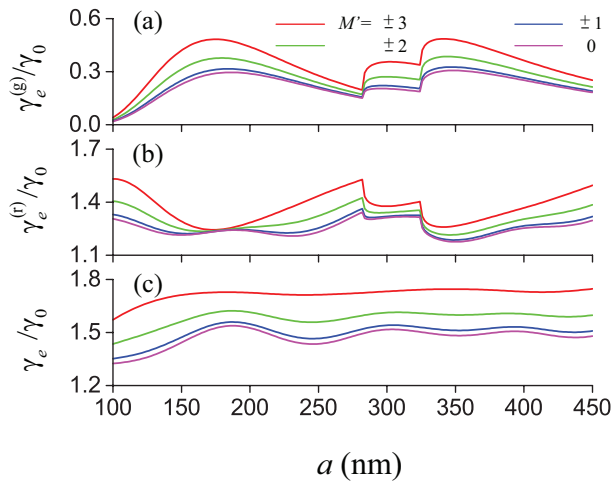}
 \end{center}
\caption{(Color online) Rates $\gamma_{e}^{(\mathrm{g})}$, $\gamma_{e}^{(\mathrm{r})}$, and $\gamma_{e}$ of spontaneous emission 
from different magnetic sublevels of the hfs level $5P_{3/2}F'=3$ into (a) guided modes, (b) radiation modes,
and (c) both types of modes as functions of the fiber radius $a$. The atom is positioned on the fiber surface. Other parameters are as for Fig.~\ref{fig2}. The rates are normalized to the free-space decay rate $\gamma_0$ of the atom.
}
\label{fig8}
\end{figure}

The rates $\gamma_{e}^{(\mathrm{g})}$, $\gamma_{e}^{(\mathrm{r})}$, and $\gamma_{e}$ of spontaneous emission from different magnetic sublevels of the hfs level $5P_{3/2}F'=3$ into guided modes, radiation modes, and  both types of modes are shown as functions of the fiber radius $a$ in Fig.~\ref{fig8}. 
We observe from the figure that, in the case where the atom is positioned on the fiber surface, the rates $\gamma_{e}^{(\mathrm{g})}$, $\gamma_{e}^{(\mathrm{r})}$, and $\gamma_{e}$ for the outermost magnetic sublevels $M'=\pm3$ are larger than for the other sublevels. The dependencies of $\gamma_{e}^{(\mathrm{g})}$ and $\gamma_{e}^{(\mathrm{r})}$ on the fiber radius $a$ are stronger than that of $\gamma_{e}$. The rates $\gamma_{e}^{(\mathrm{g})}$ and $\gamma_{e}^{(\mathrm{r})}$ undergo steep variations at the point $a\simeq 283$ nm, which corresponds to the cutoff for the TE$_{01}$ and TM$_{01}$ modes, and at the point $a\simeq 325$ nm, which corresponds to the cutoff for the HE$_{21}$ modes. Such abrupt changes are due to the changes of the mode structure at the cutoffs. It is interesting to note that the signs of the slopes of the changes of $\gamma_{e}^{(\mathrm{g})}$ and $\gamma_{e}^{(\mathrm{r})}$ at the cutoffs are opposite to each other. Due to the mutual compensation of these changes, the variations of the total decay rates $\gamma_{e}$ at the cutoffs are smooth [see Fig.~\ref{fig8}(c)].

\begin{figure}[tbh]
\begin{center}
 \includegraphics{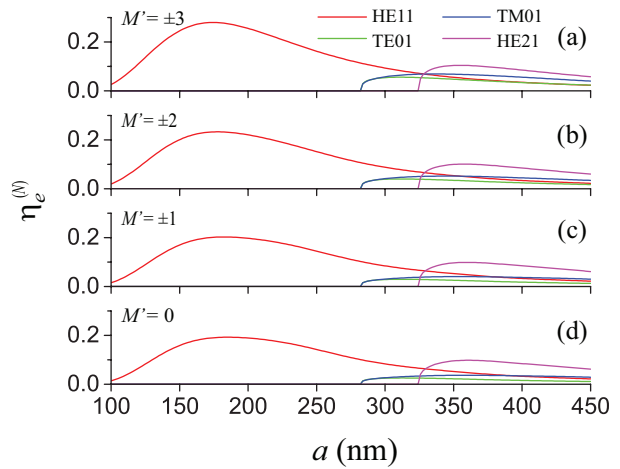}
 \end{center}
\caption{(Color online) Fractional rates $\eta_e^{(N)}=\gamma_{e}^{(N)}/\gamma_{e}$
of spontaneous emission from different magnetic sublevels of the hfs level $5P_{3/2}F'=3$
into different guided modes as functions of the fiber radius $a$. 
The atom is positioned on the fiber surface. 
Other parameters are as for Fig.~\ref{fig2}. 
}
\label{fig9}
\end{figure}

We plot in Fig.~\ref{fig9} the fractional rates $\eta_e^{(N)}=\gamma_{e}^{(N)}/\gamma_{e}$ of spontaneous emission from different magnetic sublevels of the hfs level $5P_{3/2}F'=3$ into different guided modes as functions of the fiber radius $a$. The figure shows clearly that 
the maximum value of $\eta_e^{(N)}$ for the HE$_{11}$ modes is larger than for the TE$_{01}$, TM$_{01}$, and HE$_{21}$ modes.
For a given fiber radius in the range from 330 nm to 450 nm, the value of $\eta_e^{(N)}$ for the HE$_{21}$ modes is larger than that for the other guided modes. 

\begin{figure}[tbh]
\begin{center}
 \includegraphics{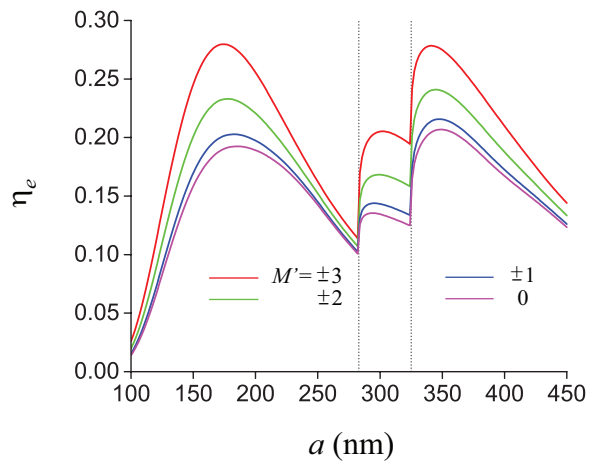}
 \end{center}
\caption{(Color online) Fractional rates $\eta_e=\gamma_{e}^{(\mathrm{g})}/\gamma_{e}$
of spontaneous emission from different magnetic sublevels of the hfs level $5P_{3/2}F'=3$ into guided modes as functions of the fiber radius $a$. 
The atom is positioned on the fiber surface. 
Other parameters are as for Fig.~\ref{fig2}. 
The vertical dotted lines indicate the positions of the cutoffs for higher-order modes.
}
\label{fig10}
\end{figure}

We show in Fig.~\ref{fig10} the fractional rates $\eta_e=\gamma_{e}^{(\mathrm{g})}/\gamma_{e}=\sum_N\eta_e^{(N)}$ of spontaneous emission from different magnetic sublevels of the hfs level $5P_{3/2}F'=3$ into all types of guided modes as functions of the fiber radius $a$. 
The figure shows that the outermost magnetic sublevels $M'=\pm3$ have the largest fractional rate. 
The fractional rates are most substantial when the fiber radius $a$ is around 180 nm and 340 nm. 
Note that, for $a\simeq 180$ nm, the fiber supports only the fundamental HE$_{11}$ modes, whereas,
for $a\simeq 340$ nm, the fiber supports not only the HE$_{11}$ modes but also the TE$_{01}$, TM$_{01}$, and HE$_{21}$ modes.

\begin{figure}[tbh]
\begin{center}
 \includegraphics{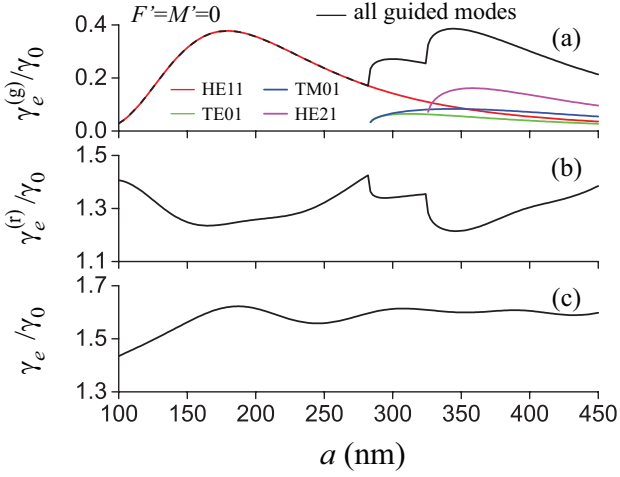}
 \end{center}
\caption{(Color online) Rates $\gamma_{e}^{(\mathrm{g})}$, $\gamma_{e}^{(\mathrm{r})}$, and $\gamma_{e}$
of spontaneous emission from the hfs level $5P_{3/2}F'=0$ into (a) guided modes, (b) radiation modes,
and (c) both types of modes as functions of the fiber radius $a$. The components $\gamma_{e}^{(N)}$ of the rate $\gamma_{e}^{(\mathrm{g})}$ are also shown in part (a). The atom is positioned on the fiber surface. Other parameters are as for Fig.~\ref{fig2}. The rates are normalized to the free-space decay rate $\gamma_0$ of the atom.
}
\label{fig11}
\end{figure}

We plot in Fig.~\ref{fig11} the spontaneous emission rates $\gamma_{e}^{(\mathrm{g})}$, $\gamma_{e}^{(\mathrm{r})}$, and $\gamma_{e}$ from the hfs level $5P_{3/2}F'=0$ into guided modes, radiation modes, and  both types of modes as functions of the fiber radius. 
As already noted in the previous subsection, the decay rate for this hfs level is equal to the average decay rate  
of an ensemble of two-level emitters with dipoles oriented randomly in space. 
Figure~\ref{fig11} shows clearly that $\gamma_{e}^{(\mathrm{g})}$ and $\gamma_{e}^{(\mathrm{r})}$ vary significantly and steeply at the cutoffs, while the variations of $\gamma_{e}$ are small and smooth.

\subsection{Dependencies of the rates on the orientation of the quantization axis}
\label{subsec:orientation}

The dipole matrix element $\mathbf{d}_{eg}$ is a vector whose spherical tensor components are specified by Eq.~\eqref{1} 
in the quantization coordinate system $\{x_Q,y_Q,z_Q\}$. It is clear that
$\mathbf{d}_{eg}$ depends on the orientation of the quantization axis $z_Q$ and so do 
the scalar product $\mathbf{d}_{eg}\cdot\mathbf{e}^{(\alpha)}$ and, hence, the spontaneous emission rate for the transition between the sublevels 
$|e\rangle$ and $|g\rangle$. 
In the previous two subsections, we have studied the case where the quantization axis $z_Q$ coincides with the fiber axis $z$.
In this subsection, we examine the dependencies of the rates on the orientation of the quantization axis.
For certainty, we assume that the atom is positioned on the axis $x$.

\begin{figure}[tbh]
\begin{center}
 \includegraphics{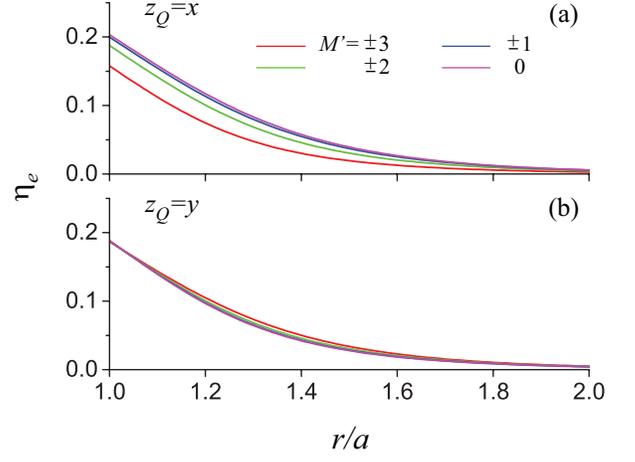}
 \end{center}
\caption{(Color online) Radial dependencies of the fractional rates $\eta_e=\gamma_{e}^{(\mathrm{g})}/\gamma_{e}$
of spontaneous emission into guided modes for
the quantization axis $z_Q=x$ (a) and $y$ (b). The atom is positioned on the axis $x$.
Other parameters are as for Fig.~\ref{fig2}.
}
\label{fig12}
\end{figure}

\begin{figure}[tbh]
\begin{center}
 \includegraphics{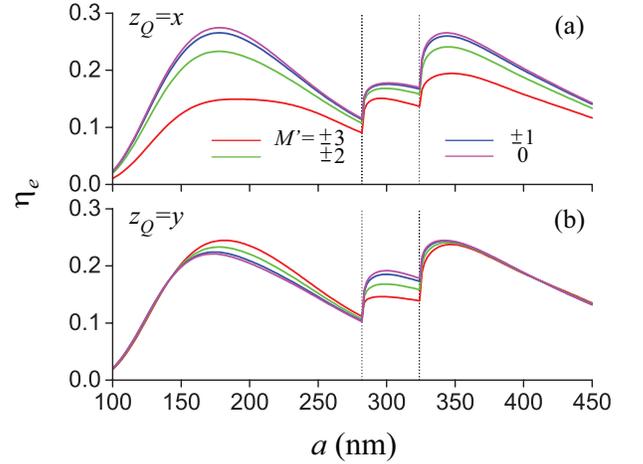}
 \end{center}
\caption{(Color online) Fractional rates $\eta_e=\gamma_{e}^{(\mathrm{g})}/\gamma_{e}$
of spontaneous emission into guided modes for
the quantization axis $z_Q=x$ (a) and $y$ (b) as functions of the fiber radius $a$. 
The atom is positioned at the crossing between the fiber surface and the axis $x$. 
Other parameters are as for Fig.~\ref{fig2}.
The vertical dotted lines indicate the positions of the cutoffs for higher-order modes.
}
\label{fig13}
\end{figure}

We plot in Figs.~\ref{fig12} and \ref{fig13} the dependencies of the fractional rates $\eta_e$
on the radial distance and the fiber radius for the quantization axis $z_Q=x$ and $y$. 
Comparison between parts (a) and (b) of these figures and between these parts and Figs.~\ref{fig5} and \ref{fig10} shows that
the rates of spontaneous emission significantly depend on the orientation of the quantization axis.
We observe that the spread of the rates with respect to the magnetic quantum number $M'$ for $z_Q=y$ [see Figs.~\ref{fig12}(b) and \ref{fig13}(b)] 
is smaller than for $z_Q=x$ [see Figs.~\ref{fig12}(a) and \ref{fig13}(a)].
    
\begin{figure}[tbh]
\begin{center}
 \includegraphics{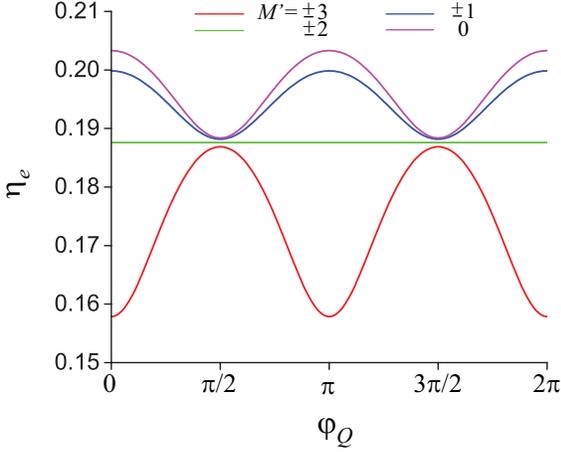}
 \end{center}
\caption{(Color online) Fractional rates $\eta_e=\gamma_{e}^{(\mathrm{g})}/\gamma_{e}$
of spontaneous emission into guided modes as functions of the azimuthal angle $\varphi_Q$ of the quantization axis $z_Q$. 
The zenithal angle of the axis $z_Q$ is $\theta_Q=\pi/2$. 
The atom is positioned at the crossing between the fiber surface and the axis $x$. Other parameters are as for Fig.~\ref{fig2}.
}
\label{fig14}
\end{figure}

\begin{figure}[tbh]
\begin{center}
 \includegraphics{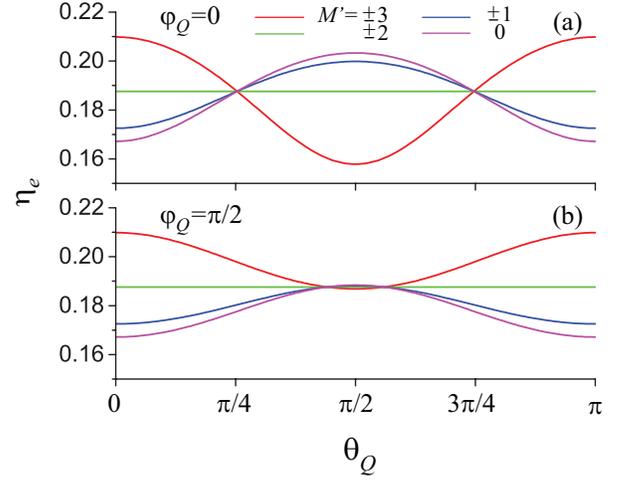}
 \end{center}
\caption{(Color online) Fractional rates $\eta_e=\gamma_{e}^{(\mathrm{g})}/\gamma_{e}$
of spontaneous emission into guided modes as functions of the zenithal angle $\theta_Q$ of the quantization axis $z_Q$. 
The azimuthal angle of the axis $z_Q$ is $\varphi_Q=0$ (a) and $\pi/2$ (b). 
The atom is positioned at the crossing between the fiber surface and the axis $x$. Other parameters are as for Fig.~\ref{fig2}.
}
\label{fig15}
\end{figure}

We plot in Figs.~\ref{fig14} and \ref{fig15} the fractional rates $\eta_e$ as functions of
the azimuthal angle $\varphi_Q$ and the zenithal angle $\theta_Q$ of the quantization axis $z_Q$. 
The figures show that the rates for the magnetic sublevels $|F'=3, M'\not=\pm2\rangle$ depend on the orientation of the quantization axis. 
It is interesting to note that the rate $\eta_e$ for the sublevels $|F'=3, M'=\pm2\rangle$ (see the green curves) does not depend on $\varphi_Q$ and $\theta_Q$. 
This independence is a consequence of the 1/2/0 ratio of the oscillatory strengths of the $\pi/\sigma_{\pm}/\sigma_{\mp}$ transitions from the magnetic sublevels $|F'=3, M'=\pm2\rangle$ \cite{coolingbook}. The symmetry properties of the profile functions with respect to opposite propagation directions and opposite phase circulation directions also play an important role.

\subsection{Directional spontaneous emission rates}
\label{subsec:asymmetry}

It has been shown in Sec. \ref{sec:spon} that, when the quantization axis $z_Q$ coincides with the fiber axis $z$ or, more generally,
lies in the meridional plane containing the position of the atom, the spontaneous emission rates $\gamma^{(Nf)}_{e}$ and $\gamma^{(\mathrm{g}f)}_{e}$ are symmetric with respect to the propagation direction $f$, that is, $\gamma^{(N+)}_{e}=\gamma^{(N-)}_{e}$ and $\gamma^{(\mathrm{g}+)}_{e}=\gamma^{(\mathrm{g}-)}_{e}$ \cite{Fam2014}. However, when the quantization axis does not lie in the meridional plane containing the position of the atom,
the decay rates $\gamma^{(Nf)}_{e}$ and $\gamma^{(\mathrm{g}f)}_{e}$ may depend on
the propagation direction $f$. In this subsection, we study the directional spontaneous emission rates for different choices of the quantization axis $z_Q$.

\begin{figure}[tbh]
\begin{center}
 \includegraphics{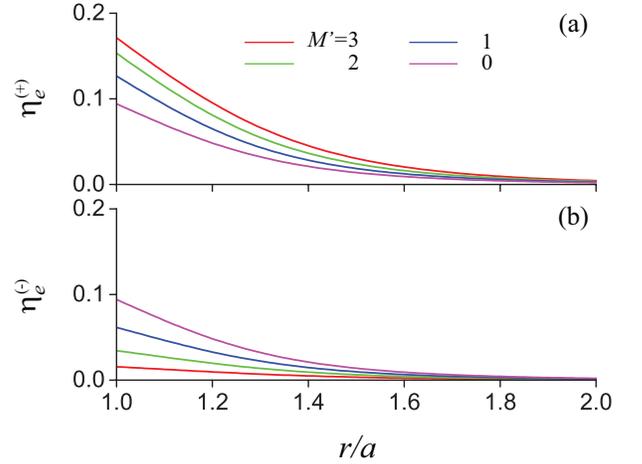}
 \end{center}
\caption{(Color online) Radial dependencies of the directional fractional rates $\eta_e^{(f)}=\gamma_{e}^{(\mathrm{g}f)}/\gamma_{e}$
of spontaneous emission into guided modes in the (a) positive and (b) negative propagation directions. 
The atom is positioned on the positive side of the axis $x$. The quantization axis $z_Q$ coincides with the axis $y$.
Other parameters are as for Fig.~\ref{fig2}.
}
\label{fig16}
\end{figure}

\begin{figure}[tbh]
\begin{center}
 \includegraphics{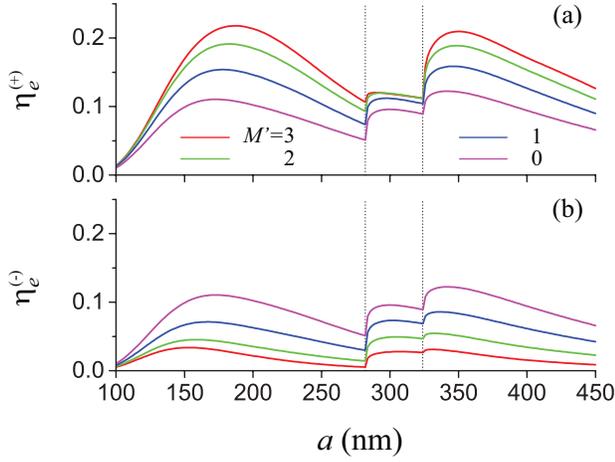}
 \end{center}
\caption{(Color online) Directional fractional rates $\eta_e^{(f)}=\gamma_{e}^{(\mathrm{g}f)}/\gamma_{e}$
of spontaneous emission into guided modes in the (a) positive and (b) negative propagation directions as functions of the fiber radius $a$. 
The atom is positioned at the point $(r=a,\varphi=0)$. The quantization axis $z_Q$ coincides with the axis $y$.
Other parameters are as for Fig.~\ref{fig2}.
The vertical dotted lines indicate the positions of the cutoffs for higher-order modes.
}
\label{fig17}
\end{figure}

The directional fractional rates $\eta_e^{(f)}=\gamma_{e}^{(\mathrm{g}f)}/\gamma_{e}$ for the positive ($f=+$) and negative ($f=-$) propagation directions 
are shown in Figs.~\ref{fig16} and \ref{fig17} as functions of the radial distance and the fiber radius. 
In the calculations of these figures, we have assumed that
the atom is positioned on the positive side of the axis $x$ and the quantization axis $z_Q$ coincides with the axis $y$.
In Figs.~\ref{fig16} and \ref{fig17}, we do not show the factor $\eta_e^{(f)}$ for $M'<0$ because it is equal to $\eta_{\bar{e}}^{(\bar{f})}$ [see Eq.~\eqref{24}]. Comparison between parts (a) and (b) of the figures shows that the directional factor $\eta_e^{(f)}$ has different values 
for different propagation directions except for the case $M'=0$ (see the magenta curves).

\begin{figure}[tbh]
\begin{center}
 \includegraphics{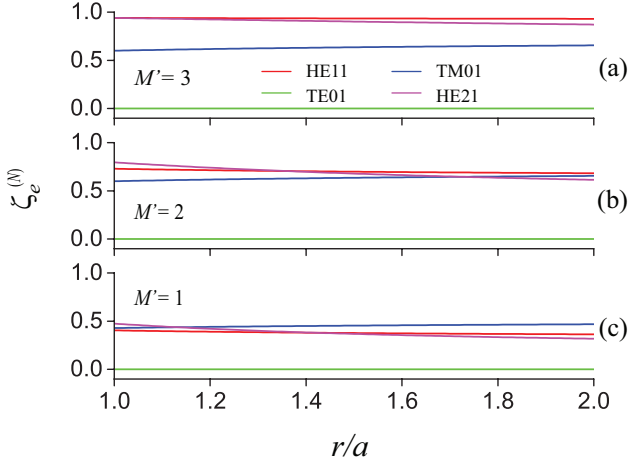}
 \end{center}
\caption{(Color online) Radial dependencies of the asymmetry factors $\zeta_e^{(N)}$
for directional spontaneous emission from different magnetic sublevels of the hfs level $5P_{3/2}F'=3$
into different guided modes. 
The atom is positioned on the positive side of the axis $x$ and the quantization axis $z_Q$ coincides with the axis $y$.
Other parameters are as for Fig.~\ref{fig2}. 
}
\label{fig18}
\end{figure}

\begin{figure}[tbh]
\begin{center}
 \includegraphics{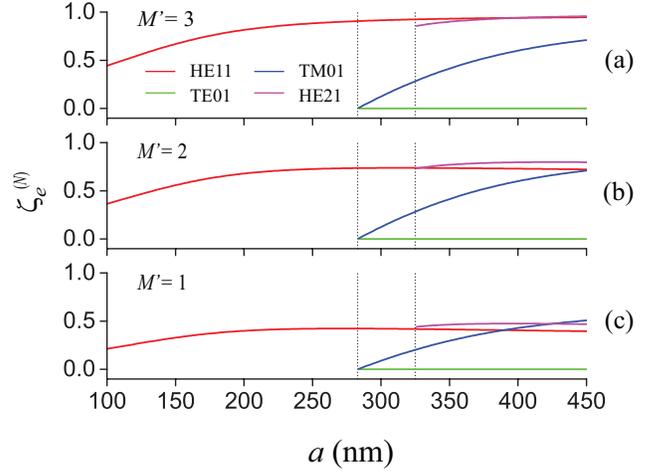}
 \end{center}
\caption{(Color online) Asymmetry factors $\zeta_e^{(N)}$
for directional spontaneous emission from different magnetic sublevels of the hfs level $5P_{3/2}F'=3$
into different guided modes as functions of the fiber radius $a$.
The atom is positioned at the point $(r=a,\varphi=0)$. The quantization axis $z_Q$ coincides with the axis $y$.
Other parameters are as for Fig.~\ref{fig2}. 
}
\label{fig19}
\end{figure}

The asymmetry between the directional rates of spontaneous emission into the positive and negative directions of the fiber axis
can be characterized by the factors
\begin{equation}\label{29}
\begin{split}
\zeta_e^{(N)}&=\frac{\gamma_{e}^{(N+)}-\gamma_{e}^{(N-)}}{\gamma_{e}^{(N+)}+\gamma_{e}^{(N-)}},\\
\zeta_e&=\frac{\gamma_{e}^{(\mathrm{g}+)}-\gamma_{e}^{(\mathrm{g}-)}}{\gamma_{e}^{(\mathrm{g}+)}+\gamma_{e}^{(\mathrm{g}-)}}.
\end{split}
\end{equation} 
We note that 
$\zeta_{\bar{e}}^{(N)}=-\zeta_e^{(N)}$ and $\zeta_{\bar{e}}=-\zeta_e$.
Hence, for the sublevel with $M'=0$, we have $\zeta_e^{(N)}=\zeta_e=0$.

We calculate numerically $\zeta_{e}^{(N)}$ for $|e\rangle$ with $M'>0$.
We show in Figs.~\ref{fig18} and \ref{fig19} the asymmetry factors $\zeta_e^{(N)}$ for directional spontaneous emission into different guided modes as functions of the radial distance and the fiber radius. 
The atom is positioned on the positive side of the axis $x$ and the quantization axis $z_Q$ coincides with the axis $y$.
We observe from Fig.~\ref{fig18} that $\zeta_e^{(N)}$ varies very slowly with increasing distance $r$.
We see from Fig.~\ref{fig19} that $\zeta_e^{(N)}$ tends to reach a stationary value when the fiber radius $a$ is large enough.
It is interesting to note that $\zeta_e^{(N)}=0$ for the TE$_{01}$ modes. The reason is that, since the longitudinal component $e^{(\mu)}_z$ of the electric part of a TE mode is zero, the profile function $\mathbf{e}^{(\mu)}$ of this mode does not depend on the propagation direction and, hence, neither does the rate for the corresponding channel of spontaneous emission.

\begin{figure}[tbh]
\begin{center}
 \includegraphics{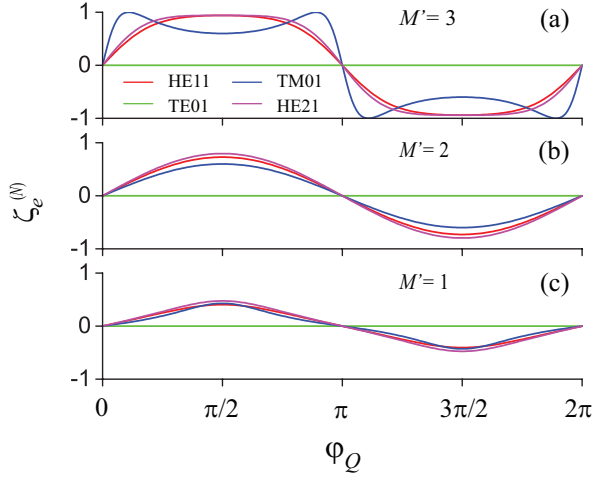}
 \end{center}
\caption{(Color online) Asymmetry factors $\zeta_e^{(N)}$
for directional spontaneous emission from different magnetic sublevels of the hfs level $5P_{3/2}F'=3$
into different guided modes as functions of the azimuthal angle $\varphi_Q$ of the quantization axis $z_Q$. 
The zenithal angle of the axis $z_Q$ is $\theta_Q=\pi/2$. 
The atom is positioned at the point $(r=a,\varphi=0)$. 
Other parameters are as for Fig.~\ref{fig2}.
}
\label{fig20}
\end{figure}

\begin{figure}[tbh]
\begin{center}
 \includegraphics{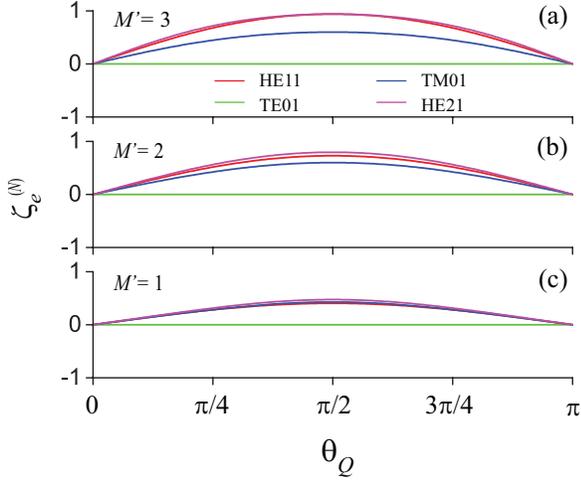}
 \end{center}
\caption{(Color online) Asymmetry factors $\zeta_e^{(N)}$
for directional spontaneous emission from different magnetic sublevels of the hfs level $5P_{3/2}F'=3$
into different guided modes as functions of the zenithal angle $\theta_Q$ of the quantization axis $z_Q$. 
The azimuthal angle of the axis $z_Q$ is $\varphi_Q=\pi/2$. 
The atom is positioned at the point $(r=a,\varphi=0)$. 
Other parameters are as for Fig.~\ref{fig2}.
}
\label{fig21}
\end{figure}

The asymmetry between the rates of spontaneous emission into the positive and negative directions of the fiber axis depends on the orientation of the quantization axis with respect to the position of the atom. We show in Figs.~\ref{fig20} and \ref{fig21} the directional asymmetry factors $\zeta_e^{(N)}$ of spontaneous emission into different guided modes as functions of 
the azimuthal angle $\varphi_Q$ and the zenithal angle $\theta_Q$ of the quantization axis $z_Q$. 
In these calculations, we assumed that the atom is positioned at the point $(r=a,\varphi=0)$. 

We observe from Fig.~\ref{fig20} and \ref{fig21} that, except for $M'=3$ and $N=\mathrm{TM}_{01}$, the absolute values of $\zeta_e^{(N)}$ are maximal when $\varphi_Q=\pi/2$ or $3\pi/2$ and $\theta_Q=\pi/2$. These angles correspond to the case where the quantization axis $z_Q$ coincides with the axis $y$. 
This axis is perpendicular to the meridional plane containing the position of the atom. 

The blue curve in Fig.~\ref{fig20}(a), which corresponds to $M'=3$, $N=\mathrm{TM}_{01}$, and $\theta_Q=\pi/2$, indicates that the absolute value of the asymmetry factor $\zeta_e^{(N)}$ is equal to 1 at four values $\varphi_Q=\varphi_0$, $\pi-\varphi_0$, $\pi+\varphi_0$,
or $2\pi-\varphi_0$, where  $\varphi_0\simeq 0.108\pi\simeq 19^\circ$. This means that the spontaneous emission from the outermost sublevel $|F'=3,M'=3\rangle$ of the hfs level $5P_{3/2}F'=3$ into the TM modes is unidirectional when the quantization axis $z_Q$ lies at an appropriate azimuthal angle $\varphi_Q$ in the fiber transverse plane $xy$. This interesting feature arises as a consequence of the properties of the cyclic transition and the TM modes. Indeed, the only allowed electric dipole transition from the sublevel $|F'=3,M'=3\rangle$ of the excited state $5P_{3/2}$ is the $\sigma_+$ transition to the sublevel $|F=2,M=2\rangle$ of the ground state $5S_{1/2}$. The dipole of this transition is coupled to the counterclockwise circular component of the projection of the electric part of the field onto the plane $x_Qy_Q$, which is perpendicular to the quantization axis $z_Q$.
When the quantization axis lies in the fiber transverse plane $xy$ and is oriented at an azimuthal angle $\varphi_Q=\varphi_0$, $\pi-\varphi_0$, $\pi+\varphi_0$,
or $2\pi-\varphi_0$, where  $\varphi_0=\arcsin(|e_z|/|e_r|)$, the polarization of the projection of the electric part of a TM mode onto the plane $x_Qy_Q$ is exactly circular at the position of the atom. The rotation direction of this polarization depends on the propagation direction $f$. 
Consequently, spontaneous emission from the sublevel $|F'=3,M'=3\rangle$ into the TM modes is unidirectional.

\section{Summary}
\label{sec:summary}

In this work, we have studied spontaneous emission from a rubidium-87 atom into the fundamental and higher-order modes of a vacuum-clad ultrathin optical fiber. We have shown that the spontaneous emission rate depends on the magnetic sublevel, the type of modes, the orientation of the quantization axis, and the fiber radius. We have found that the rate of spontaneous emission into the TE modes is always symmetric with respect to the propagation directions. 
Meanwhile, the rates of spontaneous emission into other guided modes do not depend on the propagation direction 
when the quantization axis lies in the meridional plane containing the position of the atom. Asymmetry of spontaneous emission with respect to the propagation directions may appear when the output modes are not TE modes and the quantization axis does not lie in the meridional plane containing the position of the atom. We have shown that the rate of spontaneous emission into guided modes propagating in a given direction does not change when both the propagation direction and the magnetic quantum number are reversed. This result means that asymmetry of spontaneous emission with respect to the propagation directions leads to asymmetry with respect to the magnetic quantum numbers and vice versa. For the fiber radius in the range from 330 nm to 450 nm, the spontaneous emission into the HE$_{21}$ modes is stronger than into the HE$_{11}$, TE$_{01}$, and TM$_{01}$ modes. When the quantization axis coincides with the fiber axis and the radial distance is not too large, the rates of spontaneous emission from the outermost magnetic sublevels into guided modes are larger than those from the other sublevels. At the cutoff for higher-order modes, the rates of spontaneous emission into guided and radiation modes undergo steep variations, which are caused by the changes of the mode structure. Due to the mutual compensation of these changes, the variations of the total rate of spontaneous emission into both types of modes are smooth. The total fractional rate of emission into guided modes is most substantial when the fiber radius is around 180 nm, where the fiber supports only the fundamental HE$_{11}$ modes, or 340 nm, where the fiber supports not only the HE$_{11}$ modes but also the TE$_{01}$, TM$_{01}$, and HE$_{21}$ modes. We have shown that the spontaneous emission from the upper level up the cyclic transition into the TM modes is unidirectional when the quantization axis lies at an appropriate azimuthal angle  in the fiber transverse plane. Our results lay the foundations for future research on manipulating and controlling the coupling of atoms, molecules, and dielectric particles to higher-order modes of ultrathin optical fibers.

\begin{acknowledgments}
We acknowledge support for this work from the Okinawa Institute of Science and Technology Graduate University.
S.N.C. and T.B. are grateful to JSPS for partial support from a Grant-in-Aid for Scientific Research (Grant No. 26400422). 
\end{acknowledgments}


\appendix

\section{Guided modes of a step-index fiber}
\label{sec:guided}

Consider the model of a step-index fiber that is a dielectric cylinder of radius $a$ and refractive index $n_1$ and is surrounded by an infinite background medium of refractive index $n_2$,
where $n_2<n_1$. We use the Cartesian coordinates $\{x,y,z\}$, where $z$ is the coordinate along the fiber axis.
We also use the cylindrical coordinates $\{r,\varphi,z\}$, where $r$ and $\varphi$ are the polar coordinates in the fiber transverse plane $xy$.

For a guided light field of frequency $\omega$ (free-space wavelength $\lambda=2\pi c/\omega$ and free-space wave number $k=\omega/c$), the propagation constant $\beta$ is determined by the fiber eigenvalue equation \cite{fiber books}
\begin{eqnarray}\label{a1}
\lefteqn{\bigg[\frac{J_{l}'(ha)}{haJ_{l}(ha)}
+\frac{K_{l}'(qa)}{qaK_{l}(qa)}
\bigg]\bigg[\frac{n_{1}^2J_{l}'(ha)}{haJ_{l}(ha)}
+\frac{n_{2}^2K_{l}'(qa)}{qaK_{l}(qa)}
\bigg]=}\nonumber\\&&\mbox{}\qquad\qquad\qquad\qquad\qquad
l^2\left(\frac{1}{h^2a^2}+\frac{1}{q^2a^2}\right)^2\frac{\beta^2}{k^2}.
\end{eqnarray}
Here, we have introduced the parameters $h=(n_1^2k^2-\beta^2)^{1/2}$ and $q=(\beta^2-n_2^2k^2)^{1/2}$, which characterize the scales of the spatial variations of the field inside and outside the fiber, respectively. The integer index $l=0,1,2,\dots$ is the azimuthal mode order, which determines the helical phasefront and the associated phase gradient in the fiber transverse plane. 
The notations $J_l$ and $K_l$ stand for the Bessel functions of the first kind and the modified Bessel functions of the second kind, respectively. 
The notations $J'_l(x)$ and $K'_l(x)$ stand for the derivatives of $J_l(x)$ and $K_l(x)$ with respect to the argument $x$.

For $l\geq 1$, the eigenvalue equation \eqref{a1} leads to hybrid HE and EH modes \cite{fiber books}. The eigenvalue equation is given, 
for HE modes, as
\begin{equation}\label{a2}
\frac{J_{l-1}(ha)}{haJ_{l}(ha)}=-\frac{n_{1}^2+n_{2}^2}{2n_{1}^2}\frac{K'_{l}(qa)}{qaK_{l}(qa)}+
\frac{l}{h^2a^2}-\mathcal{R}
\end{equation}
and, for EH modes, as
\begin{equation}\label{a3}
\frac{J_{l-1}(ha)}{haJ_{l}(ha)}=-\frac{n_{1}^2+n_{2}^2}{2n_{1}^2}\frac{K'_{l}(qa)}{qaK_{l}(qa)}+
\frac{l}{h^2a^2}+\mathcal{R}.
\end{equation}
Here, we have introduced the notation
\begin{equation}\label{a4}
\begin{split}
\mathcal{R}&=\bigg[\bigg(\frac{n_{1}^2-n_{2}^2}{2n_{1}^2}\bigg)^2\bigg(\frac{K'_{l}(qa)}{qaK_{l}(qa)}\bigg)^2\\
&\quad +\bigg(\frac{l\beta}{n_{1}k}\bigg)^2\bigg(\frac{1}{q^2a^2}+\frac{1}{h^2a^2}\bigg)^2\bigg]^{1/2}.
\end{split}
\end{equation}
We label HE and EH modes as HE$_{lm}$ and EH$_{lm}$, respectively, where $l=1,2,\dots$ and $m=1,2,\dots$ are the azimuthal and radial mode orders, respectively. 
Here, the radial mode order $m$ implies that the HE$_{lm}$ or EH$_{lm}$ mode is the $m$th solution to the corresponding eigenvalue equation \eqref{a2} or \eqref{a3}, respectively.

For $l=0$, the eigenvalue equation \eqref{a1} leads to TE and TM modes \cite{fiber books}. The eigenvalue equation is given, for TE modes, as
\begin{eqnarray}\label{a5}
\frac{J_{1}(ha)}{haJ_{0}(ha)}=-\frac{K_{1}(qa)}{qaK_{0}(qa)}
\end{eqnarray}
and, for TM modes, as 
\begin{eqnarray}\label{a6}
\frac{J_{1}(ha)}{haJ_{0}(ha)}=-\frac{n_2^2}{n_1^2}\frac{K_{1}(qa)}{qaK_{0}(qa)}.
\end{eqnarray}
We label TE and TM modes as TE$_{0m}$ and TM$_{0m}$, respectively, where $m=1,2,\dots$ is the radial mode order. The subscript 0 implies that the azimuthal mode order of TE and TM modes is $l=0$.
The radial mode order $m$ implies that the TE$_{0m}$ or TM$_{0m}$ mode is the $m$th solution to the corresponding eigenvalue equation \eqref{a5} or \eqref{a6}, respectively.

According to \cite{fiber books}, the fiber size parameter $V$ is defined as $V=ka\sqrt{n_1^2-n_2^2}$.
The cutoff values $V_c$ for HE$_{1m}$ modes are determined as solutions to the equation $J_1(V_c)=0$. 
For HE$_{lm}$ modes with $l=2,3,\dots$, the cutoff values are obtained as nonzero solutions to the equation $(n_1^2/n_2^2+1)(l-1)J_{l-1}(V_c)=V_cJ_l(V_c)$. The cutoff values $V_c$ for EH$_{lm}$ modes, where $l=1,2,\dots$, are determined as nonzero solutions to the equation $J_l(V_c)=0$. 
For TE$_{0m}$ and TM$_{0m}$ modes, the cutoff values $V_c$ are obtained as solutions to the equation $J_0(V_c)=0$. 

The electric component of the field can be presented in the form
\begin{equation}\label{a7}
\mathbf{E}=\frac{1}{2}\boldsymbol{\mathcal{E}}e^{-i\omega t}+\mathrm{c.c.},
\end{equation}
where $\boldsymbol{\mathcal{E}}$ is the envelope.
For a guided mode with a propagation constant $\beta$ and an azimuthal mode order $l$, we can write 
\begin{equation}\label{a8}
\boldsymbol{\mathcal{E}}=\mathbf{e}e^{i\beta z+il\varphi},
\end{equation}
where $\mathbf{e}$ is the mode profile function. In Eq.~\eqref{a8}, the parameters $\beta$ and $l$ can take not only positive but also negative values.

We decompose the vectorial function $\mathbf{e}$ into the radial, azimuthal and axial components denoted by the subscripts $r$, $\varphi$ and $z$, respectively. We summarize the expressions for the mode functions of hybrid modes, TE modes, and TM modes in the below \cite{fiber books}.

\subsection{Hybrid modes}

We consider hybrid modes $N=$ HE$_{lm}$ or EH$_{lm}$.
It is convenient to introduce the parameter
\begin{equation}\label{a9}
s=l\left(\frac{1}{h^2a^2}+\frac{1}{q^2a^2}\right)\left[\frac{J_{l}'(ha)}{haJ_{l}(ha)}
+\frac{K_{l}'(qa)}{qaK_{l}(qa)} \right]^{-1}.
\end{equation}
Then, we find, for $r<a$,
\begin{eqnarray}\label{a10}
e_{r}&=& iA\frac{\beta}{2h}[(1-s)J_{l-1}(hr)-(1+s)J_{l+1}(hr)],\nonumber\\
e_{\varphi}&=& -A\frac{\beta}{2h}[(1-s)J_{l-1}(hr)+(1+s)J_{l+1}(hr)],\nonumber\\
e_{z}&=& AJ_{l}(hr), 
\end{eqnarray}
and, for $r>a$,
\begin{eqnarray}\label{a11}
e_{r}&=& iA\frac{\beta}{2q}\frac{J_{l}(ha)}{K_{l}(qa)}[(1-s)K_{l-1}(qr)+(1+s)K_{l+1}(qr)],\nonumber\\
e_{\varphi}&=&-A\frac{\beta}{2q}\frac{J_{l}(ha)}{K_{l}(qa)}[(1-s)K_{l-1}(qr)-(1+s)K_{l+1}(qr)],\nonumber\\
e_{z}& = & A\frac{J_{l}(ha)}{K_{l}(qa)}K_{l}(qr).
\end{eqnarray}
Here, the parameter $A$ is a constant that can be determined from the propagating power of the field.
Without loss of generality, we take $A$ to be a real number.

In the cylindrical coordinates, the mode profile function of the electric component of a quasicircularly polarized hybrid mode $N$ with a propagation direction $f=\pm$  and a phase circulation direction $p=\pm$ is given by
\begin{equation}\label{a12}
\mathbf{e}^{(\omega Nfp)}=\hat{\mathbf{r}}e_r+p\hat{\boldsymbol{\varphi}}e_\varphi+f\hat{\mathbf{z}}e_z,
\end{equation}
where the mode function components $e_r$, $e_\varphi$, and $e_z$
are given by Eqs.~\eqref{a10} and \eqref{a11} for $\beta>0$ and $l>0$. 
These components depend explicitly on the azimuthal mode order $l$ and implicitly on 
the radial mode order $m$. An important property of the mode functions of hybrid modes is that the longitudinal
component $e_z$ is nonvanishing and in quadrature ($\pi/2$ out of phase) with the radial component $e_r$.
In addition, the azimuthal component $e_\varphi$ is also nonvanishing and in quadrature with the radial component $e_r$. 
We note that the full mode function of the quasicircularly polarized hybrid mode is 
$\boldsymbol{\mathcal{E}}^{(\omega Nfp)} = \mathbf{e}^{(\omega Nfp)} e^{if\beta z +ipl\varphi}$, where $\beta>0$ and $l>0$.

We have the following symmetry relations: 
\begin{eqnarray}\label{a13}
e_r^{(\omega Nfp)}&=&e_r^{(\omega N\bar{f}p)}=e_r^{(\omega Nf\bar{p})},\nonumber\\
e_{\varphi}^{(\omega Nfp)}&=&e_{\varphi}^{(\omega N\bar{f}p)}=-e_{\varphi}^{(\omega Nf\bar{p})},\nonumber\\
e_z^{(\omega Nfp)}&=&-e_z^{(\omega N\bar{f}p)}=e_z^{(\omega Nf\bar{p})},
\end{eqnarray}
and
\begin{equation}\label{a14}
e_r^{(\mu)*}=-e_r^{(\mu)},\quad
e_\varphi^{(\mu)*}=e_\varphi^{(\mu)},\quad
e_z^{(\mu)*}=e_z^{(\mu)},
\end{equation}
where $\bar{f}=-f$ and $\bar{p}=-p$. From Eqs.~\eqref{a13} and \eqref{a14}, we obtain the formulas
\begin{eqnarray}\label{a15}
e_r^{(\omega Nfp)}&=&-e_r^{(\omega N\bar{f}\bar{p})*},\nonumber\\
e_{\varphi}^{(\omega Nfp)}&=&-e_{\varphi}^{(\omega N\bar{f}\bar{p})*},\nonumber\\
e_z^{(\omega Nfp)}&=&-e_z^{(\omega N\bar{f}\bar{p})*},
\end{eqnarray}
which yield
\begin{equation}\label{a16}
\mathbf{e}^{(\omega Nfp)}=-\mathbf{e}^{(\omega N\bar{f}\bar{p})*}.
\end{equation}
Equation \eqref{a16} is a consequence of the time reversal symmetry of the field.

\subsection{TE modes}

We consider transverse electric modes $N=$ TE$_{0m}$.
For $r<a$, we have
\begin{eqnarray}\label{a17}
e_{r}&=&0,\nonumber\\
e_{\varphi}&=& i\frac{\omega\mu_{0}}{h}AJ_{1}(hr),\nonumber\\
e_{z}& = & 0. 
\end{eqnarray}
For $r>a$, we have
\begin{eqnarray}\label{a18}
e_{r}&=&0,\nonumber\\
e_{\varphi}&=&-i\frac{\omega\mu_{0}}{q}\frac{J_{0}(ha)}{K_{0}(qa)}AK_{1}(qr),\nonumber\\
e_{z}& = & 0.
\end{eqnarray}
Without loss of generality, we take $A$ to be a real number.

The mode profile function of the electric component of a TE$_{0m}$ mode with a propagation direction $f=\pm$ can be written as
\begin{equation}\label{a19}
\mathbf{e}^{(\omega \mathrm{TE}_{0m}f)}=\hat{\boldsymbol{\varphi}}e_\varphi,
\end{equation}
where the only nonzero cylindrical component $e_\varphi$ is given by the second expressions in Eqs.~\eqref{a17} and \eqref{a18}.
The mode function depends implicitly on the radial mode order $m$. 
The full mode function of the TE mode is
$\boldsymbol{\mathcal{E}}^{(\omega \mathrm{TE}_{0m}f)}= \mathbf{e}^{(\omega \mathrm{TE}_{0m}f)} e^{if\beta z}$, where $\beta>0$.

We find the relations
\begin{equation}\label{a20}
e_{\varphi}^{(\omega \mathrm{TE}_{0m}f)}=e_{\varphi}^{(\omega \mathrm{TE}_{0m}\bar{f})}=-e_{\varphi}^{(\omega \mathrm{TE}_{0m}f)*},
\end{equation}
which yield
\begin{equation}\label{a21}
\mathbf{e}^{(\omega \mathrm{TE}_{0m}f)}=-\mathbf{e}^{(\omega \mathrm{TE}_{0m}\bar{f})*}.
\end{equation}

\subsection{TM modes}

We consider transverse magnetic modes $N=$ TM$_{0m}$.
For $r<a$, we have
\begin{eqnarray}\label{a22}
e_{r}&=&-i\frac{\beta}{h}AJ_{1}(hr),\nonumber\\
e_{\varphi}&=& 0,\nonumber\\
e_{z}& = & AJ_{0}(hr). 
\end{eqnarray}
For $r>a$, we have
\begin{eqnarray}\label{a23}
e_{r}&=&i\frac{\beta}{q}\frac{J_{0}(ha)}{K_{0}(qa)}AK_{1}(qr),\nonumber\\
e_{\varphi}&=&0,\nonumber\\
e_{z}& = & \frac{J_{0}(ha)}{K_{0}(qa)}AK_{0}(qr).
\end{eqnarray}
Without loss of generality, we take $A$ to be a real number.

The mode profile function of the electric component of a TM mode with a propagation direction $f=\pm$ can be written as
\begin{equation}\label{a24}
\mathbf{e}^{(\omega \mathrm{TM}_{0m}f)}=\hat{\mathbf{r}}e_r+f\hat{\mathbf{z}}e_z,
\end{equation}
where the components $e_r$ and $e_z$ are given by the first and third expressions in Eqs.~\eqref{a22} and \eqref{a23} for $\beta>0$.
The mode function depends implicitly on the radial mode order $m$.
Like the case of hybrid modes, the longitudinal
component $e_z$ of a TM mode is nonvanishing and in quadrature ($\pi/2$ out of phase) with the radial component $e_r$.
The full mode function of the TM mode is
$\boldsymbol{\mathcal{E}}^{(\omega \mathrm{TM}_{0m}f)}= \mathbf{e}^{(\omega \mathrm{TM}_{0m}f)} e^{if\beta z}$, where $\beta>0$.

We find the relations
\begin{eqnarray}\label{a25}
e_r^{(\omega \mathrm{TM}_{0m}f)}&=&e_r^{(\omega \mathrm{TM}_{0m}\bar{f})}=-e_r^{(\omega \mathrm{TM}_{0m}f)*},\nonumber\\
e_z^{(\omega \mathrm{TM}_{0m}f)}&=&-e_z^{(\omega \mathrm{TM}_{0m}\bar{f})}=e_z^{(\omega \mathrm{TM}_{0m}f)*},
\end{eqnarray}
which yield
\begin{equation}\label{a26}
\mathbf{e}^{(\omega \mathrm{TM}_{0m}f)}=-\mathbf{e}^{(\omega \mathrm{TM}_{0m}\bar{f})*}.
\end{equation}

\section{Radiation modes of a nanofiber}
\label{sec:radiation}

We  present the electric component of the field  in the form
$\mathbf{E}=(1/2)(\boldsymbol{\mathcal{E}}e^{-i\omega t}+\mathrm{c.c.})$,
where $\boldsymbol{\mathcal{E}}$ is the envelope.
For a radiation mode with a propagation constant $\beta$ in the range $-kn_2<\beta<kn_2$ and a mode order $l=0,\pm1,\pm2,\dots$, we can write 
$\boldsymbol{\mathcal{E}}=\mathbf{e}e^{i\beta z+il\varphi}$,
where $\mathbf{e}$ is the mode profile function.
The characteristic parameters for the field in the inside and outside of the fiber are $h=\sqrt{k^2n_1^2-\beta^2}$ and $q=\sqrt{k^2n_2^2-\beta^2}$, respectively.

The mode functions of the electric parts of the radiation modes $\nu=(\omega\beta l p)$
\cite{fiber books} are given, for $r<a$, by
\begin{eqnarray}\label{q1}
e_r^{(\nu)}&=&
\frac{i}{h^2}\left[\beta hAJ'_l(hr)+il\frac{\omega\mu_0}{r}BJ_l(hr)\right],\nonumber\\ 
e_{\varphi}^{(\nu)}&=&
\frac{i}{h^2}\left[il\frac{\beta}{r}AJ_l(hr)-h\omega\mu_0BJ'_l(hr)\right],\nonumber\\
e_z^{(\nu)}&=&AJ_l(hr),
\end{eqnarray}
and, for $r>a$, by 
\begin{eqnarray}\label{q2}
e_r^{(\nu)}&=&
\frac{i}{q^2}\sum_{j=1,2}
\left[\beta q C_jH^{(j)\prime}_l(qr)+il\frac{\omega\mu_0}{r}D_jH^{(j)}_l(qr)\right],\nonumber\\
e_{\varphi}^{(\nu)}&=&
\frac{i}{q^2}\sum_{j=1,2}
\left[il\frac{\beta}{r}C_jH^{(j)}_l(qr)-q\omega\mu_0D_jH^{(j)\prime}_l(qr)\right], \nonumber\\
e_z^{(\nu)}&=&\sum_{j=1,2}C_jH_l^{(j)}(qr).
\end{eqnarray}
Here, $A$ and $B$ as well as $C_j$ and $D_j$ with $j=1,2$ are coefficients.
The coefficients $C_j$ and $D_j$ are related to the coefficients $A$ and $B$ as 
\cite{Tromborg}
\begin{eqnarray}\label{q3}
C_j&=&(-1)^{j}\frac{i\pi q^2a}{4n_2^2}(AL_j+i\mu_0cBV_j),\nonumber\\
D_j&=&(-1)^{j-1}\frac{i\pi q^2a}{4}(i\epsilon_0cAV_j-BM_j),
\end{eqnarray}
where
\begin{eqnarray}\label{q4}
V_j&=&\frac{lk\beta}{ah^2q^2}
(n_2^2-n_1^2)
J_l(ha)H_l^{(j)*}(qa),\nonumber\\
M_j&=&\frac{1}{h}J'_l(ha)H_l^{(j)*}(qa)
-\frac{1}{q}J_l(ha)H_l^{(j)*\prime}(qa),\nonumber\\
L_j&=&\frac{n_1^2}{h}J'_l(ha)H_l^{(j)*}(qa)
-\frac{n_2^2}{q}J_l(ha)H_l^{(j)*\prime}(qa).\nonumber\\
\end{eqnarray}
We specify two polarizations by choosing $B=i\eta A$ and $B=-i\eta A$ for $p=+$
and $p=-$, respectively. We take $A$ to be a real number.
The orthogonality of the modes requires
\begin{eqnarray}\label{q5}
&&\int _0^{2\pi}d\varphi\int _{0}^{\infty}n_{\mathrm{ref}}^2
\left[\mathbf{e}^{(\nu)}\mathbf{e}^{(\nu')*}\right]_{\beta=\beta',l=l'}
\;rdr \nonumber\\&&
=N_{\nu}\delta_{pp'}\delta(\omega-\omega').
\end{eqnarray}
This leads to
\begin{equation}\label{q6}
\eta=\epsilon_0c\sqrt{\frac{n_2^2|V_j|^2+|L_j|^2}{|V_j|^2+n_2^2|M_j|^2}}.
\end{equation}
The constant $N_{\nu}$ is given by 
\begin{equation}\label{q7}
N_{\nu}=\frac{8\pi \omega}{q^2}\left(n_2^2|C_j|^2+\frac{\mu_0}{\epsilon_0}|D_j|^2\right).
\end{equation}

We introduce the notations $\bar{\beta}=-\beta$, $\bar{l}=-l$, and $\bar{p}=-p$.
We find the symmetry relations 
\begin{eqnarray}\label{q8}
e_r^{(\omega\beta lp)}&=&-e_r^{(\omega\bar{\beta} l\bar{p})},\nonumber\\
e_{\varphi}^{(\omega\beta lp)}&=&-e_{\varphi}^{(\omega\bar{\beta} l\bar{p})},\nonumber\\
e_z^{(\omega\beta lp)}&=&e_z^{(\omega\bar{\beta} l\bar{p})},\nonumber\\
\end{eqnarray}
\begin{eqnarray}\label{q9}
e_{r}^{(\omega\beta lp)}&=&(-1)^l e_{r}^{(\omega\beta \bar{l}\bar{p})},\nonumber\\
e_{\varphi}^{(\omega\beta lp)}&=&(-1)^{l+1} e_{\varphi}^{(\omega\beta \bar{l}\bar{p})},\nonumber\\
e_{z}^{(\omega\beta lp)}&=&(-1)^l e_{z}^{(\omega\beta \bar{l}\bar{p})},
\end{eqnarray} 
and
\begin{equation}\label{q10}
e_r^{(\nu)*}=-e_r^{(\nu)},\quad
e_\varphi^{(\nu)*}=e_\varphi^{(\nu)},\quad
e_z^{(\nu)*}=e_z^{(\nu)},
\end{equation}
which yield
\begin{equation}\label{q11}
\mathbf{e}^{(\omega\beta lp)}=(-1)^l\mathbf{e}^{(\omega\bar{\beta} \bar{l}p)*}.
\end{equation} 

For the spherical tensor components $e_q^{(\omega,\beta, l,p)}$, with the index $q=0,\pm1$, of the radiation mode functions, we find the relations
\begin{equation}\label{q12}
e_q^{(\omega\beta l p)}=(-1)^q e_q^{(\omega\bar{\beta} l \bar{p})},
\end{equation}
\begin{equation}\label{q13}
e_{q}^{(\omega\beta l p)}=(-1)^{l+q} e^{2iq\varphi} e_{-q}^{(\omega\beta \bar{l} \bar{p})},
\end{equation}
and
\begin{equation}\label{q14}
e_{q}^{(\omega\beta l p)}=(-1)^{q} e^{2iq\varphi} e_{q}^{(\omega\beta l p)*}.
\end{equation}

\end{document}